\documentclass[a4paper,12pt]{article}
\usepackage{jheppubnew}
\usepackage{pgf,tikz,pgfplots}
\usetikzlibrary{cd}
\pgfplotsset{compat=1.15}
\usepackage{mathrsfs}
\usetikzlibrary{arrows}
\usepackage{dsfont}
\usepackage{amsfonts}
\usepackage{amsmath,amssymb}
\usepackage{physics}
\usepackage{amsthm}
\usepackage{setspace}
\usepackage{relsize}
\usepackage{todonotes}
\usepackage{xpatch}
\usepackage[none]{hyphenat}
\usetikzlibrary{matrix}

\usepackage{tikz-cd}

\xpatchcmd\section{\large}{\Large}{}{}
\xpatchcmd\subsection{\normalsize}{\large}{}{}
\xpatchcmd\subsubsection{\normalsize}{\normalsize}{}{}

\newcommand{\CD}{\mathcal{D}}

\newcommand{\CB}{\mathcal{B}}
\newcommand{\CC}{\mathcal{C}}

\newcommand{\CI}{\mathcal{I}}

\newcommand{\CS}{\mathcal{S}}
\newcommand{\CZ}{\mathcal{Z}}

\newcommand{\DZ}{\mathds{Z}}
\newcommand{\trl}{\mathbf{1}}

\newcommand{\be}{\begin{equation}}
\newcommand{\ee}{\end{equation}}
\newcommand{\bea}{\begin{eqnarray}}
\newcommand{\eea}{\end{eqnarray}}

\newcommand{\FC}{\mathfrak{C}}
\newcommand{\FD}{\mathfrak{D}}
\newcommand{\FM}{\mathfrak{M}}

\newcommand{\SC}{\mathscr{C}}
\newcommand{\SD}{\mathscr{D}}
\newcommand{\SE}{\mathscr{E}}
\newcommand{\SN}{\mathscr{N}}

 \title{On the action of non-invertible symmetries on local operators in 3+1d}
\author[1]{Pavel Putrov}
\author[1,2]{and Rajath Radhakrishnan}
\affiliation[1]{International Centre for Theoretical Physics,\\ Strada Costiera 11, Trieste 34151, Italy}
\affiliation[2]{Mathematical Institute, University of Oxford,
Andrew Wiles Building, Woodstock Road, Oxford, OX2 6GG, UK}

\abstract{Most of the known non-invertible symmetries of quantum field theories in three and four spacetime dimensions act invertibly on local operators. An exception is coset symmetries, which can be constructed from gauging a non-normal subgroup of an invertible symmetry. In this paper, we study the action of a general finite non-invertible symmetry on local operators in four dimensions. We show that non-invertible symmetries without topological line operators necessarily act invertibly on local operators. Using this result, we argue that the action of a general non-invertible symmetry in 3+1d on local operators can be decomposed into the invertible action of some operators composed with the action of a gauging interface. We use this result to study when such a symmetry is anomaly-free. We find a necessary condition for a finite non-invertible symmetry in 3+1d to be anomaly-free, and show that anomaly-free non-invertible symmetries without topological line operators are non-intrinsically non-invertible.}

\begin{document}

\maketitle

\newpage

\section{Introduction}

The realisation that all topological operators must be understood as symmetries of quantum field theories (QFTs) has led to an avalanche of research on understanding them \cite{Gaiotto:2014kfa, Schafer-Nameki:2023jdn,Brennan:2023mmt,Bhardwaj:2023kri, Luo:2023ive,Shao:2023gho}. These symmetries start showing their non-trivial nature starting from spacetime dimension two, where topological line operators can have a non-invertible action on local operators of the QFT. A quintessential and well-studied example is the Ising conformal field theory (CFT), which is one among a zoo of rational CFTs with non-invertible symmetries \cite{Frohlich:2006ch,Chang:2018iay}. 

In higher dimensions, a quantum field theory typically contains topological operators of various codimensions. For example, in 2+1d, we can have both topological line operators as well as 2-dimensional surface operators. In 3+1d, we can have topological 3-dimensional membrane operators, 2-dimensional surface operators and line operators, and so on. Naively, it would seem that this freedom in having topological operators of various codimensions will make the symmetry structure rich as we move up in spacetime dimension. While this is true, there is a sense in which topological operators become more constrained in higher dimensions. These constraints arise from the fact that topological operators cannot exist independently of each other and must satisfy various mutual consistency conditions. For example, one can take a codimension-1 topological operator and dimensionally reduce it to a higher codimension topological operator. Through this relation, the fusion rules of the higher codimension operators can impose non-trivial constraints on lower codimension operators. In fact, this has strong implications for the classification of such operators in higher dimensions. This is particularly constraining in TQFTs where the topological operators have non-trivial action on each other, which renders the classification of TQFTs relatively simpler in four spacetime dimensions and higher \cite{Lan:2018vjb,Lan:2018bui,Johnson-Freyd:2020usu,Kong:2020jne,Johnson-Freyd:2020ivj}. 

Despite these constraints, various families of non-invertible symmetries implemented by topological membrane (3-dimensional) operators are known to exist in 3+1d QFTs:
\begin{itemize}
	\item Condensation operators \cite{
Kapustin:2010if,Gaiotto:2019xmp,Roumpedakis:2022aik}. 
\item Duality defects \cite{Choi:2021kmx,Kaidi:2021xfk,Choi:2022zal}.  
\item Coset non-invertible symmetries \cite{Bhardwaj:2022lsg,Hsin:2024aqb,Hsin:2025ria}.
\end{itemize}
Among these symmetries, only coset symmetries act non-invertibly on local operators. Indeed, condensation operators by definition are created from a network of codimension $\geq 2$ operators. Since these higher codimension operators act trivially on local operators, so do the condensation membrane operators. Duality defects in 3+1d act by gauging a 1-form symmetry and relabelling of fields. Therefore, the action of this symmetry on local operators is invertible. Coset non-invertible symmetries can act non-invertibly on local operators. However, since they are constructed from gauging a subgroup of an invertible 0-form symmetry, their non-invertible action can be reduced to the action of an invertible 0-form symmetry group along with the action of a gauging interface. In summary, the above non-invertible symmetries either act invertibly on local operators, or their action can be simplified into an action of a group and a gauging interface. Given these examples, we ask the following question:

\vspace{0.2cm}
\textit{What is the action of a general non-invertible 0-form symmetry on local operators in $d>2$ spacetime dimensions?}
\vspace{0.2cm}

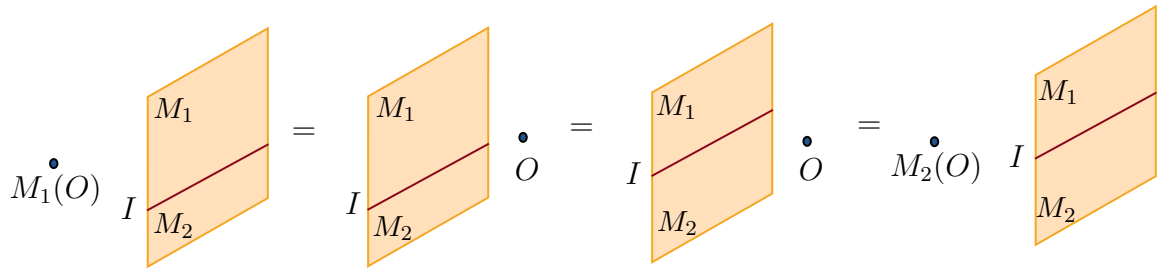
\begin{figure}[h!]
	\centering

\tikzset{every picture/.style={line width=0.75pt}} 

\begin{tikzpicture}[x=0.75pt,y=0.75pt,yscale=-1,xscale=0.9]

\draw  [color={rgb, 255:red, 245; green, 166; blue, 35 }  ,draw opacity=1 ][fill={rgb, 255:red, 255; green, 224; blue, 187 }  ,fill opacity=1 ] (374.14,157.99) -- (374.16,72.57) -- (440.77,38.07) -- (440.76,123.49) -- cycle ;
\draw [color={rgb, 255:red, 139; green, 6; blue, 24 }  ,draw opacity=1 ]   (373.78,114.69) -- (441.14,81.37) ;
\draw  [color={rgb, 255:red, 0; green, 0; blue, 0 }  ,draw opacity=1 ][fill={rgb, 255:red, 32; green, 83; blue, 143 }  ,fill opacity=1 ] (458.04,97.21) .. controls (458.04,95.96) and (458.95,94.96) .. (460.08,94.96) .. controls (461.21,94.96) and (462.12,95.96) .. (462.12,97.21) .. controls (462.12,98.45) and (461.21,99.46) .. (460.08,99.46) .. controls (458.95,99.46) and (458.04,98.45) .. (458.04,97.21) -- cycle ;
\draw  [color={rgb, 255:red, 245; green, 166; blue, 35 }  ,draw opacity=1 ][fill={rgb, 255:red, 255; green, 224; blue, 187 }  ,fill opacity=1 ] (586.85,149.27) -- (586.86,63.85) -- (653.47,29.34) -- (653.46,114.76) -- cycle ;
\draw [color={rgb, 255:red, 139; green, 6; blue, 24 }  ,draw opacity=1 ]   (586.48,105.96) -- (653.84,72.65) ;
\draw  [color={rgb, 255:red, 0; green, 0; blue, 0 }  ,draw opacity=1 ][fill={rgb, 255:red, 32; green, 83; blue, 143 }  ,fill opacity=1 ] (528.7,97.38) .. controls (528.7,96.14) and (529.61,95.13) .. (530.74,95.13) .. controls (531.87,95.13) and (532.78,96.14) .. (532.78,97.38) .. controls (532.78,98.63) and (531.87,99.64) .. (530.74,99.64) .. controls (529.61,99.64) and (528.7,98.63) .. (528.7,97.38) -- cycle ;
\draw  [color={rgb, 255:red, 245; green, 166; blue, 35 }  ,draw opacity=1 ][fill={rgb, 255:red, 255; green, 224; blue, 187 }  ,fill opacity=1 ] (216.61,159.99) -- (216.62,74.57) -- (283.23,40.07) -- (283.22,125.49) -- cycle ;
\draw [color={rgb, 255:red, 139; green, 6; blue, 24 }  ,draw opacity=1 ]   (216.24,131.69) -- (283.6,98.37) ;
\draw  [color={rgb, 255:red, 0; green, 0; blue, 0 }  ,draw opacity=1 ][fill={rgb, 255:red, 32; green, 83; blue, 143 }  ,fill opacity=1 ] (300.5,95.21) .. controls (300.5,93.96) and (301.41,92.96) .. (302.54,92.96) .. controls (303.67,92.96) and (304.58,93.96) .. (304.58,95.21) .. controls (304.58,96.45) and (303.67,97.46) .. (302.54,97.46) .. controls (301.41,97.46) and (300.5,96.45) .. (300.5,95.21) -- cycle ;
\draw  [color={rgb, 255:red, 245; green, 166; blue, 35 }  ,draw opacity=1 ][fill={rgb, 255:red, 255; green, 224; blue, 187 }  ,fill opacity=1 ] (94.35,160.27) -- (94.36,74.85) -- (160.97,40.34) -- (160.96,125.76) -- cycle ;
\draw [color={rgb, 255:red, 139; green, 6; blue, 24 }  ,draw opacity=1 ]   (93.98,131.96) -- (161.34,98.65) ;
\draw  [color={rgb, 255:red, 0; green, 0; blue, 0 }  ,draw opacity=1 ][fill={rgb, 255:red, 32; green, 83; blue, 143 }  ,fill opacity=1 ] (40.2,108.38) .. controls (40.2,107.14) and (41.11,106.13) .. (42.24,106.13) .. controls (43.37,106.13) and (44.28,107.14) .. (44.28,108.38) .. controls (44.28,109.63) and (43.37,110.64) .. (42.24,110.64) .. controls (41.11,110.64) and (40.2,109.63) .. (40.2,108.38) -- cycle ;

\draw (453.82,104.29) node [anchor=north west][inner sep=0.75pt]    {$O$};
\draw (374.59,72.51) node [anchor=north west][inner sep=0.75pt]  [font=\small]  {$M_{1}$};
\draw (375.32,130.74) node [anchor=north west][inner sep=0.75pt]  [font=\small]  {$M_{2}$};
\draw (357.67,106.35) node [anchor=north west][inner sep=0.75pt]    {$I$};
\draw (486.09,85.73) node [anchor=north west][inner sep=0.75pt]    {$=$};
\draw (504.5,101.08) node [anchor=north west][inner sep=0.75pt]    {$M_{2}( O)$};
\draw (586.43,64.78) node [anchor=north west][inner sep=0.75pt]  [font=\small]  {$M_{1}$};
\draw (585.16,124.01) node [anchor=north west][inner sep=0.75pt]  [font=\small]  {$M_{2}$};
\draw (570.37,97.63) node [anchor=north west][inner sep=0.75pt]    {$I$};
\draw (296.28,104.29) node [anchor=north west][inner sep=0.75pt]    {$O$};
\draw (219.05,73.51) node [anchor=north west][inner sep=0.75pt]  [font=\small]  {$M_{1}$};
\draw (217.43,133.09) node [anchor=north west][inner sep=0.75pt]  [font=\small]  {$M_{2}$};
\draw (202.13,121.35) node [anchor=north west][inner sep=0.75pt]    {$I$};
\draw (326.55,86.73) node [anchor=north west][inner sep=0.75pt]    {$=$};
\draw (16,112.08) node [anchor=north west][inner sep=0.75pt]    {$M_{1}( O)$};
\draw (95.93,73.78) node [anchor=north west][inner sep=0.75pt]  [font=\small]  {$M_{1}$};
\draw (95.66,132.01) node [anchor=north west][inner sep=0.75pt]  [font=\small]  {$M_{2}$};
\draw (77.87,123.63) node [anchor=north west][inner sep=0.75pt]    {$I$};
\draw (172.55,88.73) node [anchor=north west][inner sep=0.75pt]    {$=$};

\end{tikzpicture}
\caption{The action of a topological codimension 1 operator on a local operator $O$ is defined by sliding the topological operator past $O$. If two codimension 1 operators $M_1$ and $M_2$ are connected by a topological interface $I$, then $I$ can be freely deformed so as to implement either the action of $M_1$ or that of $M_2$ on $O$. Consequently, the actions of $M_1$ and $M_2$ on $O$ must coincide.}
	\label{fig: action of M1 S M2}
\end{figure}

To answer this question, we first note that finite non-invertible symmetries of a QFT in $d+1$ spacetime dimensions are implemented by topological operators of codimension-1 all the way to codimension-0. Even though this collection of operators can be very complicated in general, only the codimension-1 operators can non-trivially act on local operators. Moreover, the distinct codimension-1 operators do not all have distinct actions on the local operators. Consider the action of a topological codimension-1 operator on a local operator $O$ defined by sliding the topological operator past $O$. Two codimension-1 operators $M_1$ and $M_2$ have equal action on local operators if they are connected by a topological codimension-2 interface. This is because the topological interface can be freely moved, and we can make either $M_1$ or $M_2$ pass through the local operator $O$ (see Fig. \ref{fig: action of M1 S M2}). As a special case, consider condensation membrane operators. They admit at least one topological boundary condition. In other words, a condensation operator admits a topological interface with the identity codimension-1 operator and hence must act trivially on local operators. 

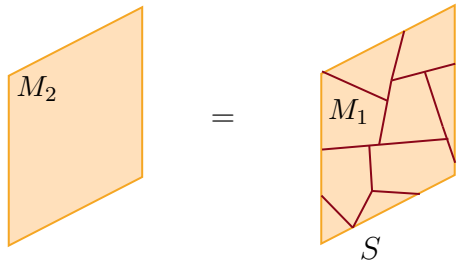
\begin{figure}[h!]
	\centering

\tikzset{every picture/.style={line width=0.75pt}} 

\begin{tikzpicture}[x=0.75pt,y=0.75pt,yscale=-1,xscale=1]

\draw  [color={rgb, 255:red, 245; green, 166; blue, 35 }  ,draw opacity=1 ][fill={rgb, 255:red, 255; green, 224; blue, 187 }  ,fill opacity=1 ] (201.61,167.99) -- (201.62,82.57) -- (268.23,48.07) -- (268.22,133.49) -- cycle ;
\draw  [color={rgb, 255:red, 245; green, 166; blue, 35 }  ,draw opacity=1 ][fill={rgb, 255:red, 255; green, 224; blue, 187 }  ,fill opacity=1 ] (357.61,166.99) -- (357.62,81.57) -- (424.23,47.07) -- (424.22,132.49) -- cycle ;
\draw [color={rgb, 255:red, 139; green, 6; blue, 24 }  ,draw opacity=1 ]   (399.01,60) -- (392.6,84.98) ;
\draw [color={rgb, 255:red, 139; green, 6; blue, 24 }  ,draw opacity=1 ]   (424.5,76.5) -- (392.6,84.98) ;
\draw [color={rgb, 255:red, 139; green, 6; blue, 24 }  ,draw opacity=1 ]   (392.6,84.98) -- (386.54,116.98) ;
\draw [color={rgb, 255:red, 139; green, 6; blue, 24 }  ,draw opacity=1 ]   (357.99,80.36) -- (390.58,95.11) ;
\draw [color={rgb, 255:red, 139; green, 6; blue, 24 }  ,draw opacity=1 ]   (409.34,80.74) -- (418.24,108.11) ;
\draw [color={rgb, 255:red, 139; green, 6; blue, 24 }  ,draw opacity=1 ]   (418.24,108.11) -- (424.48,126.33) ;
\draw [color={rgb, 255:red, 139; green, 6; blue, 24 }  ,draw opacity=1 ]   (382.98,140.5) -- (406.84,142.04) ;
\draw [color={rgb, 255:red, 139; green, 6; blue, 24 }  ,draw opacity=1 ]   (358,119.68) -- (421.09,114.28) ;
\draw [color={rgb, 255:red, 139; green, 6; blue, 24 }  ,draw opacity=1 ]   (381.56,117.37) -- (382.98,140.5) ;
\draw [color={rgb, 255:red, 139; green, 6; blue, 24 }  ,draw opacity=1 ]   (382.98,140.5) -- (373.37,159) ;
\draw [color={rgb, 255:red, 139; green, 6; blue, 24 }  ,draw opacity=1 ]   (357.71,142.81) -- (373.37,159) ;

\draw (204.05,81.51) node [anchor=north west][inner sep=0.75pt]  [font=\small]  {$M_{2}$};
\draw (359.62,92.97) node [anchor=north west][inner sep=0.75pt]  [font=\small]  {$M_{1}$};
\draw (301,100.4) node [anchor=north west][inner sep=0.75pt]    {$=$};
\draw (375.37,162.4) node [anchor=north west][inner sep=0.75pt]    {$S$};

\end{tikzpicture}
	\caption{If the topological operators $M_1$ and $M_1$ are related by a topological interface $I$, then $M_2$ can be constructed by putting a network of $S:=I^{\dagger}\times I$ lines on it. Since $S$ is a codimension-2 operator, it does not act on local operators.}
	\label{fig:M2 as condensation of M1}
\end{figure}

Another way to understand this is to note that if $M_1$ and $M_2$ are connected by a topological interface, $M_2$ is obtained from $M_1$ by gauging some operators on $M_1$. Therefore, $M_2$ can be understood as $M_1$ with a network of operators on it (see Fig. \ref{fig:M2 as condensation of M1}). However, since the network of operators on $M_1$ are codimension 2 or higher, they do not act on local operators. Therefore, the action of $M_1$ and $M_2$ on local operators must be identical. 

In $d+1$ spacetime dimensions, one can also consider the action of $M_1$ defined on $S^{d}$ on a local operator $O(x)$, where $x$ is a local inside the sphere. We can introduce a local patch of $M_2$ on $S^{d}$ along with the interface $I$. Sweeping the interface $I$ over the $S^d$ gives the operator $M_2$. Hence, codimension-1 operators which admit a topological interface between them are identified on $S^{d-1}$, and therefore have the same action on local operators.\footnote{To capture the full action of a non-invertible symmetry on local operators, one must consider the action of the symmetry on twisted-sector local operators defined at the end of line operators. In this case, the action of a codimension one operator $M$ defined on $S^{d-1}$ depends on the choice of a junction with the line operator hosting the twisted-sector local operator. Two codimension one operators admitting a topological interface between them have identical action on twisted-sector local operators up to a change in the choice of this junction. For an explicit discussion of this in 2+1d, see \cite[Sec. 3]{Bartsch:2023wvv}. (See also \cite{Bhardwaj:2023ayw}.)} 

 The existence of a topological interface defines an equivalence relation on the topological operators. As far as the action on local operators is concerned, we only have to consider the equivalence classes of codimension-1 operators.
In this paper, we will show that in the absence of topological line operators, the codimension-1 operators act invertibly on local operators. We show this by arguing that the equivalence classes of codimension-1 operators can be equipped with a group-like fusion. For general non-invertible symmetries, described by a collection of topological operators, we can have non-trivial topological line operators. In this case, the equivalence classes of codimension-1 operators do not have a group-like fusion. However, in four spacetime dimensions and higher, such a symmetry can be reduced to a symmetry without topological line operators by gauging away all the lines. Using this relation, we will show that one can decompose the action of a codimension-1 operator on a local operator in terms of certain codimension-1 operators which act invertibly on local operators, along with the action of a gauging interface. As we will review, this is precisely how coset non-invertible symmetries act on local operators. 

When the action of the non-invertible symmetry on local operators is invertible, the local operators are in some $G$-representation for some finite group $G$.  One may wonder whether there is an effective 't Hooft anomaly for $G$ which obstructs the existence of a trivially gapped phase with this symmetry. We will make this intuition precise by showing that non-invertible symmetries without topological line operators are anomaly-free only if there exists a gauging that renders the codimension-1 operators invertible up to condensation. Moreover, this invertible symmetry is precisely $G$ and must be non-anomalous. A special case of this result recovers the statement that non-anomalous duality defects in 3+1d are group theoretical \cite{Sun:2023xxv}. For general non-invertible symmetries involving topological line operators, we find a constraint on the structure of the topological line operators for the symmetry to be anomaly-free. In this case, we find that there is a topological manipulation which renders a subset of the topological operators invertible. 

The plan of the paper is as follows. In preparation for the analysis in 3+1d, we start with Section \ref{sec: 2+1d action of non-invertibles} in which we review that in the absence of topological line operators, topological surface operators in 2+1d must be invertible. We then study the action of general non-invertible surface operators on local operators. In Section \ref{sec:Action on local operators in 3+1d}, we study the action of 3-dimensional topological membrane operators on local operators. We show that, in the absence of topological line operators, the set of equivalence classes of membrane operators can be associated with a group-like fusion. We use this result to write an expression for the action of a general non-invertible symmetry in 3+1d on local operators. In Section \ref{sec:anomaly-free 3+1d}, we study anomaly-free non-invertible symmetries. We start with anomaly-free non-invertible symmetries without topological line operators and show that the topological operators implementing them can be rendered invertible up to condensations. We also derive an explicit necessary condition for a general non-invertible symmetry to be anomaly-free. Finally, in Section \ref{sec:conclusion}, we describe the generalization of our results to higher fusion categories containing fermionic line operators as well as to higher dimensions and conclude with some future directions. 

\text{Note:} Throughout this paper, we assume that the only topological local operator is the identity operator. We only consider non-invertible symmetries of bosonic QFTs implemented by topological operators where every topological operator is a finite sum of a collection of simple topological operators. We assume that there are no topological fermionic line operators in the collection of topological operators describing a non-invertible symmetry in 3+1d. This is an assumption made to simplify our arguments, and they can be generalizaed to the case with fermionic topological lines as discussed in the conclusions.  

\section{Action on local operators in 2+1d}

\label{sec: 2+1d action of non-invertibles}

In 2+1d, non-invertible symmetries are implemented by topological surface and line operators. Mathematically, they form a fusion 2-category $\FC$ whose objects describe surface operators, 1-morphisms describe line operators at the junction of two surface operators, and 2-morphisms describe local operators at the junction of two line operators \cite{Bhardwaj:2022yxj,douglas2018fusion}. In particular, the genuine line operators can be understood as the possible junctions of the trivial surface operator with itself. The genuine line operators will be denoted as $\Omega \FC$. They form a braided fusion category. 

In 2+1d, only the surface operators can act non-trivially on local operators. Moreover, as described in the introduction, surface operators related by a topological junction have the same action on local operators. Consider the equivalence class of surface operators where two surface operators are considered equivalent if they are related by a topological junction. Mathematically, these are the Schur components of the fusion 2-category denoted as $\pi_0(\FC)$. In the following discussion, we will first consider the case of non-invertible symmetries where the only topological line operator is the identity line before moving on to the general case.

In principle, the results obtained in this section can be derived from the classification of fusion 2-categories \cite{Bhardwaj:2024qiv,Decoppet:2024htz,Bhardwaj:2025piv}. However, such a classification is not available in 3+1d. Therefore, our approach will be different and will be presented in such a way that it can be generalized to higher dimensions. 

\subsection{Non-invertible symmetries without line operators}

\label{sec: 2+1d action of non-invertibles without lines}

Let us assume that the non-invertible symmetry whose topological operators are described by $\FC$ does not contain any non-trivial topological line operators. In other words, we assume that $\Omega\FC\cong \text{Vec}$, where Vec is the category of vector spaces. The action of surface operators on local operators only depends on the equivalence classes in $\pi_0(\FC)$. In the next subsection, we will use the SymTFT of $\FC$ to show that the set $\pi_0(\FC)$ can be equipped with a group-like fusion rule . This then implies that the local operators are in a representation of this group. Finally, we will arrive at the same result using dimensional reduction of topological operators.

\subsubsection{Group-like fusion on $\pi_0(\FC)$ from 3+1d SymTFT}

\begin{figure}[h!]
	\centering

\tikzset{every picture/.style={line width=0.75pt}} 

\begin{tikzpicture}[x=0.75pt,y=0.75pt,yscale=-1,xscale=1]

\draw  [fill={rgb, 255:red, 255; green, 224; blue, 187 }  ,fill opacity=1 ] (306.5,87) -- (379.5,87) -- (379.5,204) -- (306.5,204) -- cycle ;
\draw  [color={rgb, 255:red, 0; green, 0; blue, 0 }  ,draw opacity=1 ][fill={rgb, 255:red, 74; green, 74; blue, 74 }  ,fill opacity=0.38 ] (260.15,217.1) -- (260.17,99.24) -- (337,78) -- (336.98,195.86) -- cycle ;
\draw [color={rgb, 255:red, 139; green, 6; blue, 24 }  ,draw opacity=1 ][fill={rgb, 255:red, 245; green, 166; blue, 35 }  ,fill opacity=1 ][line width=0.75]    (306.5,86) -- (306.5,204) ;
\draw  [dash pattern={on 4.5pt off 4.5pt}]  (175.23,100.21) -- (398.5,97) ;
\draw  [dash pattern={on 4.5pt off 4.5pt}]  (217.23,80.21) -- (440.5,77) ;
\draw  [dash pattern={on 4.5pt off 4.5pt}]  (173.23,219.21) -- (396.5,216) ;
\draw  [dash pattern={on 4.5pt off 4.5pt}]  (212.23,198.21) -- (435.5,195) ;

\draw (262.17,102.64) node [anchor=north west][inner sep=0.75pt]  [font=\small]  {$\CB_{\FC}$};
\draw (171,103.4) node [anchor=north west][inner sep=0.75pt]    {$\CZ(\FC)$};
\draw (433.5,102.4) node [anchor=north west][inner sep=0.75pt]    {$G$-SPT};
\draw (353,98.4) node [anchor=north west][inner sep=0.75pt]    {$M_{g}$};
\draw (292,139.4) node [anchor=north west][inner sep=0.75pt]    {$S$};

\end{tikzpicture}
	\caption{The gapped boundary $\CB_{\FC}$ can be understood as an interface between the TQFT $\CZ(\FC)$ and a $G$-SPT obtained from gauging the $\Omega \CZ(\FC)\cong \text{Rep}(G)$ symmetry. A surface operator $S$ on $\CB_{\FC}$ is attached to a membrane operator $M_g$ in the $G$-SPT implementing the dual 0-form symmetry. We have suppressed one dimension of all operators in the diagram.}
	\label{SymTFT SPT interface}
\end{figure}
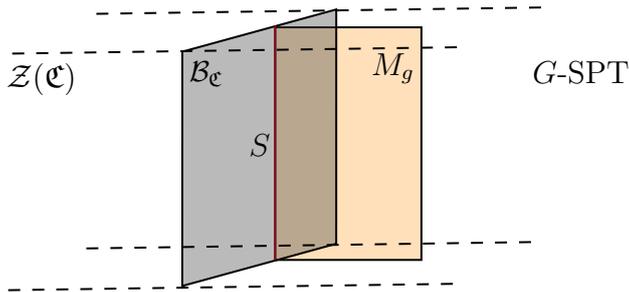

Consider the 3+1d SymTFT $\CZ(\FC)$ of the non-invertible symmetry $\FC$ \cite{Freed:2012bs,Gaiotto:2020iye,Kong:2020cie,Apruzzi:2021nmk,Freed:2022qnc,Chatterjee:2022kxb,Kaidi:2022cpf,Kaidi:2023maf}.\footnote{Mathematically, $\CZ(\FC)$ is the Drinfeld centre of the fusion 2-category $\FC$ \cite[Section 3]{baez1996higher}. (See also \cite{CRANS1998183})} By construction, $\CZ(\FC)$ has a canonical 2+1d gapped boundary condition, say $\CB_{\FC}$ on which the topological operators form the category $\FC$. Topological line operators braid trivially in $3+1$d. Therefore, the line operators in $\CZ(\FC)$ are Wilson lines for a gauge group $G$.\footnote{Here we assume that there are no fermionic line operators. Our analysis can be readily generalized to this case.} Specifically, using Deligne's theorem \cite{deligne2002categories}, we have
\be
\Omega \CZ(\FC)\cong \text{Rep}(G)~.
\ee
By assumption, the category $\FC$ does not contain any non-trivial topological line operators. Therefore, all the line operators in Rep$(G)$ must become trivial on the boundary $\CB_{\FC}$. In other words, the boundary condition $\CB_{\FC}$ is created by gauging all the Rep$(G)$ line operators. On gauging the line operators, we get a theory with a dual 0-form $G$ symmetry implemented by topological membrane operators $M_g,\;g\in G$. Moreover, on gauging Rep$(G)$, all the surface operators which braid non-trivially with the line operators get confined. In other words, they become topological 2d boundary conditions of the $M_g$ operators. In fact, since $\CZ(\FC)$ is a TQFT, all surface operators up to condensations must braid non-trivially with Rep$(G)$. Therefore, all surface operators get confined under gauging Rep$(G)$. The resulting 3+1d TQFT is a $G$-SPT. The boundary $\CB_{\FC}$ can be understood as a topological interface between the 3+1d TQFTs $\CZ(\FC)$ and a $G$-SPT. The topological surface operators on $\CB_{\FC}$ (and therefore in $\FC$) are precisely the twisted-sector boundary surface operators of the $M_g$ membrane operators in the $G$-SPT (see Fig. \ref{SymTFT SPT interface}).

Let $\FD_g$ be the 2-category of surface and line operators at the boundary of the codimension-1 operator $M_g$. Then, it is clear that there are no topological interfaces between the surface operators in $\FD_g$ and $\FD_h$ for $g\neq h$. Moreover, $\FD_g$ contains a unique simple object. To see this, note that in a $G$-SPT, each $M_g$ has a unique boundary condition. This is because the 2-categories $\FD_g$ provide a $G$-extension of the trivial fusion 2-category 2Vec \cite{Decoppet:2024moc,Decoppet:2025eic}: 
\be
\FD:=\bigoplus_{g\in G} \FD_g~, 
\ee
where $\CD_e=\text{2Vec}$. In particular, this extended category is a fusion 2-category in which the fusion rules of surface operators respect the grading by $G$. We have
\be
S_g \times \overline{S_g} \in \FD_{e}~,
\ee
for some simple surface operator $S_g \in \FD_g$ and $\overline{S_g}$ is the orientation reversal of $S_g$. Since the only simple surface operator in $\FD_e$ is the identity operator $\mathds{1}$, we have $S_g \times \overline{S_g}=\mathds{1}$. Therefore, $S_g$ is invertible. Now, suppose $S_g$ and $S'_g$ are two simple objects in $\FD_g$. We have
\be
S_g \times \overline{S'_g} =\mathds{1}~.
\ee
Since these surface operators are invertible, we have $S_g = S'_g$. Therefore, each $\FD_g$ contains a unique simple object. This shows that the surface operators of $\FC$ are precisely $S_g$ for $g \in G$. Moreover, their fusion respects the group multiplication rules
\be
S_g \times S_h =S_{gh}~.
\ee
Therefore, in the absence of line operators, the surface operators in $\FC$ are all invertible, and therefore act invertibly on the local operators. This is consistent with the fact that $\CZ(\FC)$ is a $G$-Dijkgraaf-Witten (DW) theory. The boundary condition obtained from gauging the Rep$(G)$ 2-form symmetry of $\CZ(\FC)$ must be 2Vec$_G^{\pi}$ for some 4-cocycle $\pi \in Z^4(G,U(1))$.\footnote{Note that the SymTFT of 2Vec$_G^{\pi} \boxtimes \FM$ for $\FM$ a modular tensor category (MTC) is also a $G$-DW theory. However, in our case the boundary we are interested in does not contain any line operators. Therefore, it must be described by  2Vec$_G^{\pi}$.}

\subsubsection{Dimensional reduction of topological surface operators}

\label{sec:2+1d dimensional reduction}

In the previous section, we used the 3+1d SymTFT of $\FC$ to argue that if $\FC$ does not contain line operators, then all the surface operators in it must be invertible. In this section, we will prove the same result using the dimensional reduction of surface operators into line operators. For fusion 2-categories, this proof was given in \cite{Johnson-Freyd:2020ivj}, following the results of \cite{Lan:2018vjb,Kong:2020jne}. A physical proof of the same statement was also explained in \cite{Buican:2023bzl}. Here we reproduce this argument. A generalization of this argument will be used in higher dimensions. 

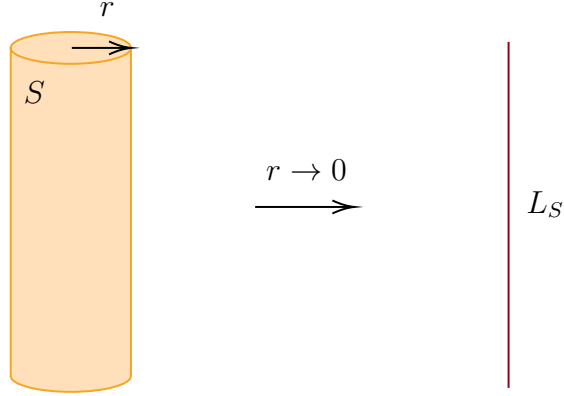
\begin{figure}
	\centering

\tikzset{every picture/.style={line width=0.75pt}} 

\begin{tikzpicture}[x=0.75pt,y=0.75pt,yscale=-1,xscale=1]

\draw  [color={rgb, 255:red, 245; green, 166; blue, 35 }  ,draw opacity=1 ][fill={rgb, 255:red, 255; green, 224; blue, 187 }  ,fill opacity=1 ] (262,60) -- (262,225) .. controls (262,229.42) and (248.57,233) .. (232,233) .. controls (215.43,233) and (202,229.42) .. (202,225) -- (202,60) .. controls (202,55.58) and (215.43,52) .. (232,52) .. controls (248.57,52) and (262,55.58) .. (262,60) .. controls (262,64.42) and (248.57,68) .. (232,68) .. controls (215.43,68) and (202,64.42) .. (202,60) ;
\draw    (232.5,60) -- (260,60) ;
\draw [shift={(262,60)}, rotate = 180] [color={rgb, 255:red, 0; green, 0; blue, 0 }  ][line width=0.75]    (10.93,-3.29) .. controls (6.95,-1.4) and (3.31,-0.3) .. (0,0) .. controls (3.31,0.3) and (6.95,1.4) .. (10.93,3.29)   ;
\draw    (324,140) -- (371.5,140) ;
\draw [shift={(373.5,140)}, rotate = 180] [color={rgb, 255:red, 0; green, 0; blue, 0 }  ][line width=0.75]    (10.93,-3.29) .. controls (6.95,-1.4) and (3.31,-0.3) .. (0,0) .. controls (3.31,0.3) and (6.95,1.4) .. (10.93,3.29)   ;
\draw [color={rgb, 255:red, 139; green, 6; blue, 24 }  ,draw opacity=1 ]   (450.5,57) -- (450.5,231) ;

\draw (245,35.4) node [anchor=north west][inner sep=0.75pt]    {$r$};
\draw (207,75.4) node [anchor=north west][inner sep=0.75pt]    {$S$};
\draw (328,114.4) node [anchor=north west][inner sep=0.75pt]    {$r\rightarrow 0$};
\draw (458,130.4) node [anchor=north west][inner sep=0.75pt]    {$L_{S}$};

\end{tikzpicture}
	\caption{Consider a surface operator $S\in \FC$ on $S^{1} \times \mathds{R}$. At the limit where the radius of $S^1$ goes to zero, we get a line operator $L_S$. $L_S$ must be topological as the operator $S$ is topological.}
	\label{fig: dimensional reduction of surface operator}
\end{figure}

Let $S \in \FC$ be a surface operator. Putting it on $S^1 \times \mathds{R}$ and shrinking the $S^1$ produces a line operator in $\Omega\FC$ (see Fig. \ref{fig: dimensional reduction of surface operator}). Let $L_S$  be this line operator. $L_S$ is, in general, a non-simple line operator. It can be understood as the action of the surface operator $S$ on the identity line operator $\trl$. The action of $S$ on $\trl$ clearly contains $\trl$.
We have
\be
L_S=n_S \trl + \dots~.
\ee
where $n_S$ is some non-zero integer. Moreover, if $S$ is a simple surface operator $n_S=1$. This is because $n_S$ is the dimension of the vector space of topological point operators at the junction of the $\trl$ line with the surface $S$. Therefore, if $n_S>1$, then there must be multiple topological local operators on the surface $S$, contradicting the assumption that $S$ is simple. If $\FC$ does not contain any non-trivial line operators, then $L_S=\trl$ for all $S$. 

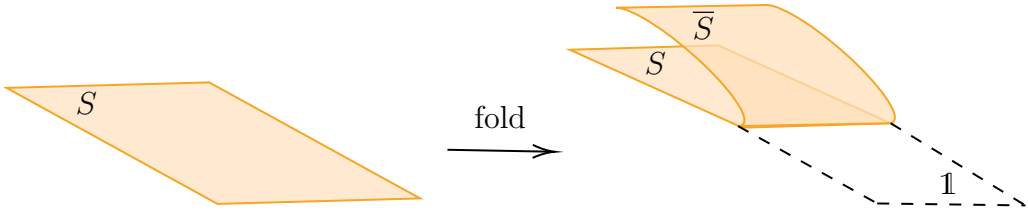
\begin{figure}[h!]
	\centering

\tikzset{every picture/.style={line width=0.75pt}} 

\begin{tikzpicture}[x=0.75pt,y=0.75pt,yscale=-1,xscale=1]

\draw  [color={rgb, 255:red, 245; green, 166; blue, 35 }  ,draw opacity=1 ][fill={rgb, 255:red, 255; green, 224; blue, 187 }  ,fill opacity=0.67 ] (357.04,96.69) -- (431.02,94.46) -- (517.22,133.66) -- (443.24,135.89) -- cycle ;
\draw  [color={rgb, 255:red, 245; green, 166; blue, 35 }  ,draw opacity=1 ][fill={rgb, 255:red, 255; green, 224; blue, 187 }  ,fill opacity=0.74 ] (516.13,133.69) -- (441.03,135.07) .. controls (448.93,134.92) and (441.7,121.48) .. (424.87,105.05) .. controls (408.04,88.62) and (387.99,75.41) .. (380.09,75.56) -- (455.19,74.18) .. controls (463.09,74.04) and (483.14,87.24) .. (499.97,103.67) .. controls (516.79,120.11) and (524.03,133.54) .. (516.13,133.69) -- cycle ;
\draw  [color={rgb, 255:red, 245; green, 166; blue, 35 }  ,draw opacity=1 ][fill={rgb, 255:red, 255; green, 224; blue, 187 }  ,fill opacity=0.67 ] (75.92,115.87) -- (177.06,113.1) -- (282.34,171.46) -- (181.2,174.23) -- cycle ;
\draw  [dash pattern={on 4.5pt off 4.5pt}]  (441.03,135.07) -- (510.5,174) ;
\draw  [dash pattern={on 4.5pt off 4.5pt}]  (517.22,133.66) -- (584.5,175) ;
\draw  [dash pattern={on 4.5pt off 4.5pt}]  (510.5,174) -- (584.5,175) ;
\draw    (296,147) -- (347.5,147.96) ;
\draw [shift={(349.5,148)}, rotate = 181.07] [color={rgb, 255:red, 0; green, 0; blue, 0 }  ][line width=0.75]    (10.93,-3.29) .. controls (6.95,-1.4) and (3.31,-0.3) .. (0,0) .. controls (3.31,0.3) and (6.95,1.4) .. (10.93,3.29)   ;

\draw (308,123.4) node [anchor=north west][inner sep=0.75pt]    {fold};
\draw (109,116.4) node [anchor=north west][inner sep=0.75pt]    {$S$};
\draw (393,96.4) node [anchor=north west][inner sep=0.75pt]    {$S$};
\draw (417,75.4) node [anchor=north west][inner sep=0.75pt]    {$\overline{S}$};
\draw (540,157.4) node [anchor=north west][inner sep=0.75pt]    {$\mathds{1}$};

\end{tikzpicture}
	\caption{Folding a topological surface operator $S$ gives an interface between the surface operator $S\times \overline{S}$ and the identity surface operator $\mathds{1}$. This shows that the outcome of the fusion rule $S\times \overline{S}$ must contain at least one condensation surface operator.}
	\label{fig: folding surface operator}
\end{figure}

Consider folding the surface operator $S$ to get the surface operator $S\times \overline{S}$ (see Fig. \ref{fig: folding surface operator}). Clearly, $S\times \overline{S}$ admits an interface with the identity surface operator. Therefore, the outcomes of the fusion $S\times \overline{S}$ must contain a condensation surface operator. We have
\be
\label{eq: fusion of S and its dual}
S \times \bar S = C + \dots~,
\ee
where $C$ is a condensation surface operator. 

Consider the surface operators $S$ and $\overline{S}$ defined on $S^1 \times \mathds{R}$ such that the $S^1$ support of both surface operators are concentric. Moreover, suppose the radius of the $S^{1}$ support of $\overline{S}$, denoted by $r_{\overline{S}}$, is larger than $r_{S}$ similarly defined. We can fuse the operators $S$ and $\overline{S}$ on $S^1 \times \mathds{R}$ by taking the limit of $r_{\overline{S}}-r_{S} \to 0$, and then shrink the $S^1$ support of the resulting operators to get the line operator $L_{S \times \overline{S}}$. Alternatively, we can shrink the $S^1$ support of $S$ to get the line operator $L_{S}$. Now, shrinking the $S^1$ support of $\overline{S}$ gives the action of $\overline{S}$ on the line operator $L_S$. We have 
\be
L_{S \times \overline{S}}= \overline{S} (L_S)~,
\ee
where $\overline{S} (L_S)$ is the topological line operator obtained from the action of $\overline{S}$ on $L_S$.
Since $L_S=\trl$ for all $S$, we have 
\be
L_{S \times \overline{S}}= \overline{S} (\trl)= \trl~.
\ee
Therefore, $S \times \overline{S}$ must be a simple surface operator. Using \eqref{eq: fusion of S and its dual}, we  have 
\be
S\times \overline{S}=C~,
\ee
Since there are no non-trivial line operators in $\FC$, $C$ must be the identity surface operator. This shows that all surface operators in $\FC$ are invertible.\footnote{Similar results do not hold when dimensionally reducing line operators. Indeed, even in the absence of non-trivial topological local operators, topological line operators can be non-invertible. In this case, the dimensional reduction of a topological line operator on $S^1$ gives a real number times the identity local operator. This real number is called the quantum dimension of the line, and it satisfies the fusion rules. This is unlike the dimensional reduction of lower codimension operators to line operators which must give a linear combination of simple line operators with non-negative integer coefficients.} Therefore, the action of these surface operators on local operators is invertible. 

\subsection{Action of general finite non-invertible symmetries}

\label{sec: 2+1d action of non-invertibles with lines}

In the previous section, we showed that non-invertible symmetries described by fusion 2-categories $\FC$ such that $\Omega \FC\cong \text{Vec}$ must act invertibly on local operators. In this section, we will study the action of a general fusion 2-category on local operators. We start with a discussion of coset non-invertible symmetries before considering the general case. 

\subsubsection{Coset non-invertible symmetries}

\begin{figure}[h!]
	\centering

\tikzset{every picture/.style={line width=0.75pt}} 

\begin{tikzpicture}[x=0.75pt,y=0.75pt,yscale=-1,xscale=0.8]

\draw  [color={rgb, 255:red, 0; green, 0; blue, 0 }  ,draw opacity=1 ][fill={rgb, 255:red, 74; green, 74; blue, 74 }  ,fill opacity=0.38 ] (121.16,169.08) -- (121.18,62.18) -- (185.22,42.91) -- (185.2,149.81) -- cycle ;
\draw  [dash pattern={on 4.5pt off 4.5pt}]  (50.36,63.06) -- (236.49,60.14) ;
\draw  [dash pattern={on 4.5pt off 4.5pt}]  (85.37,44.92) -- (271.5,42) ;
\draw  [dash pattern={on 4.5pt off 4.5pt}]  (46.69,170) -- (232.82,167.08) ;
\draw  [dash pattern={on 4.5pt off 4.5pt}]  (81.2,151.95) -- (267.33,149.04) ;
\draw  [color={rgb, 255:red, 0; green, 0; blue, 0 }  ,draw opacity=1 ][fill={rgb, 255:red, 74; green, 74; blue, 74 }  ,fill opacity=0.38 ] (404.16,167.73) -- (404.18,60.83) -- (468.22,41.56) -- (468.2,148.46) -- cycle ;
\draw  [dash pattern={on 4.5pt off 4.5pt}]  (333.36,61.71) -- (606.5,60) ;
\draw  [dash pattern={on 4.5pt off 4.5pt}]  (368.37,43.57) -- (639.5,41) ;
\draw  [dash pattern={on 4.5pt off 4.5pt}]  (332.5,168) -- (606.5,168) ;
\draw  [dash pattern={on 4.5pt off 4.5pt}]  (364.2,150.6) -- (632.5,150) ;
\draw  [color={rgb, 255:red, 0; green, 0; blue, 0 }  ,draw opacity=1 ][fill={rgb, 255:red, 74; green, 74; blue, 74 }  ,fill opacity=0.38 ] (515.16,167.73) -- (515.18,60.83) -- (579.22,41.56) -- (579.2,148.46) -- cycle ;

\draw (126.84,67.37) node [anchor=north west][inner sep=0.75pt]  [font=\small]  {$\CI$};
\draw (0.84,65.24) node [anchor=north west][inner sep=0.75pt]    {$\small \FC(G,H,\pi,\omega)$};
\draw (409.84,66.02) node [anchor=north west][inner sep=0.75pt]  [font=\small]  {$\CI$};
\draw (285.84,63.89) node [anchor=north west][inner sep=0.75pt]    {$\small \FC(G,H,\pi,\omega)$};
\draw (521.84,66.02) node [anchor=north west][inner sep=0.75pt]  [font=\small]  {$\CI^{\dagger}$};
\draw (200.84,64.24) node [anchor=north west][inner sep=0.75pt]    {2Vec$_G^{\pi}$};
\draw (455.93,64.25) node [anchor=north west][inner sep=0.75pt]    {2Vec$_G^{\pi}$};
\draw (593.93,65.25) node [anchor=north west][inner sep=0.75pt]    {$\FC(G,H,\pi,\omega)$};

\end{tikzpicture}
	\caption{Left: The gauging interface $I$ separating $\FC(G,H,\pi,\omega)$ and 2Vec$_G^{\pi}$. Right: The fusion of interfaces $\CI^{\dagger} \times \CI$  is a surface operators in $\FC(G,H,\pi,\omega)$.}
	\label{fig: gauging interface in 2+1d}
\end{figure}
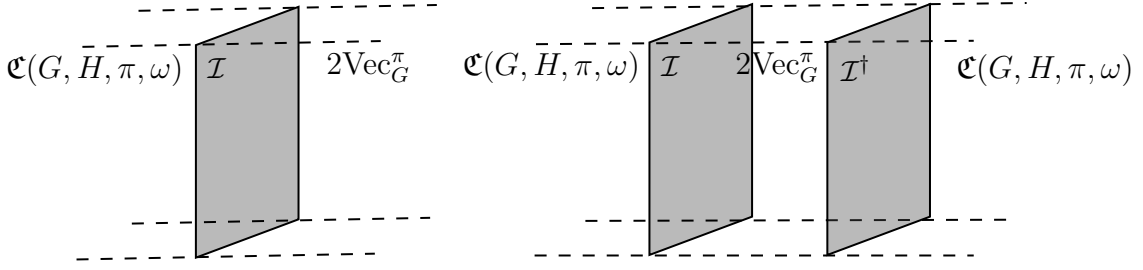

Consider the fusion 2-category 2Vec$_G^{\pi}$ where $\pi \in Z^4(G,U(1))$ which describes a 0-form $G$ symmetry with anomaly $\pi$. We will denote the surface operators implementing this symmetry as $S_g,~g\in G$. Consider a non-anomalous subgroup $H$. We have $\pi|_H=d \omega$ for some 3-cochain $\omega$. On gauging this subgroup, we get a fusion 2-category 
\be
\FC(G,H,\pi,\omega)~.
\ee
This is the coset symmetry describing a symmetry implemented by the double cosets $H\backslash G/H$. When $H$ is a non-normal subgroup of $G$, this is a non-invertible symmetry. We have 
\be
\Omega \FC(G,H,\pi,\omega) \cong \text{Rep}(H)~,
\ee
which is the dual 1-form symmetry. Let $S$ be a simple object in $\FC(G,H,\pi,\omega)$. To study the action of $S$ on local operators, let us introduce the topological gauging interface $\CI$ between $\FC(G,H,\pi,\omega)$ and 2Vec$_G^{\pi}$ obtained from gauging the 1-form symmetry Rep$(G)$ (see Fig. \ref{fig: gauging interface in 2+1d}). $\CI$ can be understood as a map form the operators in $\FC(G,H,\pi,\omega)$ to 2Vec$_G^{\pi}$ where $\CI(S)$ is the image of $S$ under gauging Rep$(H)$. Now, the composition $\CI^{\dagger} \times \CI$ is a condensation surface operator in $\FC(G,H,\pi,\omega)$ \cite{Buican:2023bzl,KNBalasubramanian:2025vum}. This is because 2Vec$_G^{\pi}$ is obtained from inserting a network of the Rep$(H)$ operators in $\FC(G,H,\pi,\omega)$. On fusing $\CI$ with $\CI^{\dagger}$, this network becomes a network of operators on the surface of $\CI^{\dagger} \times \CI$, which is precisely the surface operator obtained from higher-gauging the 1-form symmetry Rep$(H)$ on a 2-manifold. Since $\CI^{\dagger} \times \CI$ is a condensation defect, it cannot act on local operators. Now, consider the action of $S$ on a local operator. The action of $S$ is the same as the action of the operator 
\be
(\CI^{\dagger} \times \CI) \times S \times (\CI^{\dagger} \times \CI)= \CI^{\dagger} \times (\CI \times S \times \CI^{\dagger}) \times \CI=\CI^{\dagger} \times S' \times \CI~,
\ee
where $S':=\CI \times S \times \CI^{\dagger}$ is a surface operator in 2Vec$_G^{\pi}$. Therefore, the action of $S$ on a local operator, say $O$, is given by
\be
S(O)=(\CI^{\dagger} \times S' \times \CI) (O)= \CI^{\dagger}(S^{'} (O))~,
\ee
where in the second equality we used the fact that the interface $\CI$ does not act on local operators as it acts by gauging a 1-form symmetry. $S'$ is not necessarily a simple object in 2Vec$_G^{\pi}$. In general, it decomposes into a direct sum of invertible surfaces. Suppose 
\be
S'= \sum_{g \in G} n_g S_g~,
\ee
where $n_g$ is a non-negative integer. The action of $S$ on a local operator is given by 
\be
\label{eq: action on coset symmetries}
S(O)= \CI^{\dagger}\bigg (\sum_{g \in G} n_g S_g (O)\bigg ) ~.
\ee
This shows that the action of $S$ is determined by the action of the invertible surfaces $S_g$ on local operators as well as the action of the interface $\CI^{\dagger}$. The interface $\CI^{\dagger}$ acts as the image of an operator under gauging the 0-form symmetry $H$. If the operator $\sum_{g \in G} n_g S_g (O)$, is charged under the $H$ 0-form symmetry, then it is confined by the $\CI^{\dagger}$ action. In other words, such operators become non-genuine point operators. Clearly, the action of $S$ on local operators is non-invertible, in general. 

The above discussion implies that the action of a general coset symmetry operator on local operators can be decomposed into the action of a 0-form symmetry and a gauging interface. This is of course not surprising given that the coset symmetry is obtained from 2Vec$_G^{\pi}$ via gauging. In the next section, we will find that the action of a general non-invertible symmetry on local operators can also be decomposed into an action by invertible surface operators and the action of a gauging interface. 
\subsubsection{The general case}

In general, the non-invertible symmetry $\FC$ contains some line operators $\Omega \FC=\CC$. Let $\CC_0$ be the subcategory of line operators which braid trivially all line operators in $\CC$. $\CC_0$ is the collection of transparent line operators in $\CC$ as they cannot be detected by braiding with other line operators. We have $\CC_0 \cong \text{Rep}(G)$ for some finite group $G$. Consider the topological interface $\CI$ between $\FC$ and another fusion 2-category $\FD$ obtained from gauging the $\CC_0$ 1-form symmetry. Following the discussion in the previous subsection, the action of a general surface operator $S$ on a local operator $O$ can be written as
\be 
S(O)=(\CI^{\dagger} \times S' \times \CI) (O)= \CI^{\dagger}(S^{'} (O))~,
\ee
where $S':=\CI  \times S \times \CI^{\dagger}$ is a surface operator in $\FD$. By construction, there are no transparent line operators in $\Omega \FD$. Therefore, $\Omega \FD$ is a modular tensor category (MTC). Consider a simple surface operator $Q$ in $\FD$. Suppose its action on the identity line operator $\trl$ satisfies
\be 
\label{eq:trivial action on the trivial line}
Q(\trl)= n_Q \trl~~ \text{ and }~~ \overline{Q}(\trl)=n_{\overline{Q}}\trl~, 
\ee
for some non-negative integers $n_Q,n_{\overline{Q}}$. Since $Q$ is a simple surface operator, $n_Q=n_{\overline{Q}}=1$. Therefore, the line operator $L_Q$ introduced in Section \ref{sec:2+1d dimensional reduction} for this surface operator is $\trl$. Following the argument there, we find that 
\be
Q \times \overline{Q}=C~,
\ee
where $C$ is a condensation surface operator. It is shown in \cite{Buican:2023bzl} that under the action of a non-invertible condensation surface $C$, either some non-trivial line operator is mapped to the trivial line $\trl$ and/or the trivial line is mapped to a non-trivial line operator. In other words, either $C$ or $\overline{ C}$ must act non-trivially on $\trl$. Given that $Q$ and $\overline{Q}$ act trivially on the identity line operator, $C$ must be an invertible condensation surface. This implies that $Q$ is invertible. This argument shows that all simple surface operators in $\FD$ which act trivially on the line operators in the MTC $\Omega \FD$  are invertible. 
 
On the other hand, suppose $Q$ acts non-trivially on $\trl$. Then it must be a non-invertible symmetry of the MTC $\Omega\FD$. All faithful symmetries of an MTC are implemented by condensation surface operators \cite{Kapustin:2010if,Fuchs:2012dt}. Therefore, we have 
\be 
Q = P \times C~,
\ee
 where $P$ is some invertible surface operator acting trivially on $\Omega \FD$ and $C$ is a condensation surface operator. This discussion implies that the surface operator $S'$ defined above is of the form
\be
S'= \sum_{i} n_i (P_i \times C_i)~.
\ee 
where $n_i$ are non-negative integers, and the sum is over some simple surface operators in $\FD$ of the form $P_i \times C_i$, where $P_i$ are invertible, and $C_i$ are condensation surfaces. The action of $S'$ on local operators is completely determined by how the invertible surface operators $P_i$ act on the local operators. Therefore, we have 
\be 
S(O)= \CI^{\dagger}(S^{'} (O))= \CI^{\dagger}\bigg (\sum_{i} n_i (P_i \times C_i) (O)\bigg )= \CI^{\dagger}\bigg (\sum_{i} n_i P_i(O)\bigg )~.
\ee
The action of a general non-invertible symmetry $S$ is determined by the action of the invertible surface operators $P_i$ along with the gauging interface $\CI^{\dagger}$. 

\subsection{Anomaly-free non-invertible symmetries in 2+1d}

\label{sec:anomaly-free 2+1d}

A non-invertible symmetry described by a fusion 2-category $\FC$ is called anomaly-free if it admits a trivially gapped phase. Mathematically, this is the requirement that $\FC$ admits a fibre 2-functor \cite{Bhardwaj:2017xup,Thorngren:2019iar}. Clearly, if the topological line operators in $\Omega \FC$ braid non-trivially with each other, then $\FC$ cannot admit a trivially gapped phase as the non-trivial braiding is an anomaly for the 1-form symmetry implemented by the line operators. Therefore, an anomaly-free fusion 2-category $\FC$ satisfies\footnote{We assume that there is no transparent fermionic line operators, but the discussion can be readily generalized to this case.}
\be
\Omega \FC = \text{Rep}(H)
\ee
for some finite group $H$. This shows that anomaly-free non-invertible symmetries are all coset symmetries. Such fusion 2-categories can be obtained from gauging a subgroup of an invertible 0-form symmetry $G$ with some anomaly $\pi \in Z^4(G,U(1))$.\footnote{Such fusion 2-categories are called group-theoretical. Note that not all group-theoretical fusion 2-categories are anomaly-free. The necessary and sufficient conditions for this are derived in \cite{Decoppet:2023bay}.} 

The above discussion implies that anomaly-free non-invertible symmetries in 2+1d are not intrinsically non-invertible, as they can be rendered invertible by a topological manipulation \cite{Kaidi:2022uux}. This is unlike 1+1d, where some anomaly-free non-invertible symmetries cannot be obtained from finite groups \cite{SaavedraRivano1972,deligne2002categories,Lu:2025yru}. In Sec. \ref{sec:anomaly-free 3+1d}, we will study anomaly-free non-invertible symmetries in 3+1d.  

\section{Action on local operators in 3+1d}

\label{sec:Action on local operators in 3+1d}

In 3+1d, a non-invertible symmetry is described by a fusion 3-category $\SC$ of topological operators \cite{Bhardwaj:2022yxj,Johnson-Freyd:2020usu,Bhardwaj:2024xcx}. These capture the data of topological membrane, surface and line operators. The topological surface and line operators form a braided fusion 2-category denoted as $\Omega \SC$. The topological line operators form a symmetric fusion category denoted as $\Omega^2 \SC$. Since topological lines braid trivially in 3+1d, we must have 
\be
\Omega^2 \SC\cong \text{Rep}(H)~,
\ee
for some finite group $H$. The existence of a topological interface defines an equivalence relation on the membrane operators. Such equivalence classes of membrane operators are the Schur components of $\SC$, which form a set $\pi_0(\SC)$. Similarly, the equivalence classes of surface operators defined analogously will be denoted as $\pi_0(\Omega \SC)$. 

The only operators which can act non-trivially on local operators are topological membrane operators supported on three-dimensional submanifolds of the spacetime. In the following subsections, we will find constraints on the fusion rules of topological membrane operators and use them to determine their action on local operators. 

\subsection{Non-invertible symmetries without line operators}

Consider a non-invertible symmetry of a 3+1d QFT described by a fusion 3-category $\SC$. We will assume that there are no topological line operators. That is
\be
\Omega^2 \SC\cong \text{Vec}~.
\ee
Unlike in 2+1d, this assumption does not render all other topological operators invertible. In fact, an important class of non-invertible symmetries satisfying this constraint are duality defects \cite{Choi:2021kmx,Kaidi:2021xfk}. We will show that the set of equivalence classes of membrane operators $\pi_0(\SC)$ can be equipped with a group structure for some finite group $G$ and that the local operators are in a $G$-representation. We will show this using the 4+1d SymTFT $\CZ(\SC)$ of $\SC$. We will also give an independent argument for invertible action on local operators using dimensional reduction of membrane operators on $S^2 \times \mathds{R}$. 

\label{sec: 3+1d action of non-inv without lines}

\subsubsection{Group-like fusion on $\pi_0(\SC)$ from 4+1d SymTFT}

Consider the SymTFT $\CZ(\SC)$ of $\SC$ which admits a canonical boundary condition $\CB_{\SC}$ on which the topological operators form the fusion 3-category $\SC$. Consider the topological line operators in $\CZ(\SC)$ described by a symmetric fusion category
\be
\Omega^2 \CZ(\SC)\cong \text{Rep}(G)~,
\ee
for some finite group $G$. Let us understand the structure of the membrane operators in the SymTFT. To that end, note that on gauging the Rep$(G)$ symmetry, we get a new TQFT described by a fusion 3-category $\SD$. Namely, the category $\SD$ governs the genuine membrane operators in the gauged theory as well as its surface and line operators. There is a dual 0-form $G$ symmetry acting on $\SD$ implemented by 4-dimensional topological operators, say $K_g$, $g\in G$. Because $\CZ(\SC)$ is a TQFT, all non-condensation membrane operators in it braid non-trivially with at least one topological line operator in $\Omega^2\CZ(\SC)$. Therefore, on gauging $\Omega^2\CZ(\SC)$, all non-condensation membrane operators get confined. In other words, they become boundary conditions of $K_g$. The boundary conditions of $K_g$ give a $G$-extension of $\SD$ given by 
\be
\SD_G:=\bigoplus_{g\in G} \SD_g~,
\ee
where $\SD_g$ is the 3-category of topological operators living at the boundary of $K_g$ and $\SD_e:=\SD$, where $e$ is the identity element of the group $G$. The fusion rules of the membrane operators in $\SD_G$ are $G$-graded. The TQFT $\CZ(\SC)$ is obtained from gauging the 0-form symmetry $G$ of $\SD$. All non-condensation membrane operators in $\CZ(\SD)$ arise from boundary conditions of the $K_g$ operators.

 Consider the bulk-to-boundary map $F_{\SC}: \CZ(\SC) \to \SC$ which specifies the map of bulk operators in $\CZ(\SC)$ to the operators on the boundary $\CB_{\SC}$. Since $\SC$ does not contain any non-trivial topological line operators, we have
\be
F_{\SC}(\Omega^2 \CZ(\SC))=F_{\SC}(\text{Rep}(G))\cong \text{Vec}~.
\ee
In other words, all line operators in $\Omega \CZ(\SC)$ can end on the boundary $\CB_{\SC}$. The boundary condition $\CB_{\SC}$ can be understood as an interface between the SymTFT $\CZ(\SC)$ and an SPT created from gauging a symmetry of $\CZ(\SC)$ specified by a Lagrangian algebra $L_{\SC}$. Since all line operators in $\Omega \CZ(\SC)$ can end on $\CB_{\SC}$, they are trivialized in the gauged theory. Therefore, they are part of the Lagrangian algebra $L_{\SC}$. From our discussion above, we find that all membrane operators in $\CZ(\SC)$ are confined on the boundary $\CB_{\SC}$. Therefore, the resulting SPT must have a dual 0-form symmetry implemented by 4-dimensional operators $K_g$. The membrane operators in $\SC$ are determined by the fusion 3-category $\SD_g$ of boundary conditions of the $K_g$ operators. 

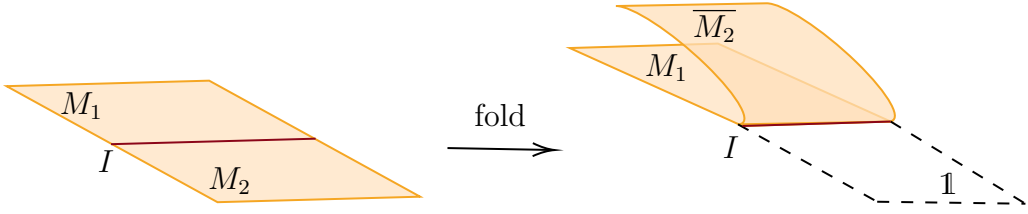
\begin{figure}[h!]
 	\centering

\tikzset{every picture/.style={line width=0.75pt}} 

\begin{tikzpicture}[x=0.75pt,y=0.75pt,yscale=-1,xscale=1]

\draw  [color={rgb, 255:red, 245; green, 166; blue, 35 }  ,draw opacity=1 ][fill={rgb, 255:red, 255; green, 224; blue, 187 }  ,fill opacity=0.67 ] (357.04,96.69) -- (431.02,94.46) -- (517.22,133.66) -- (443.24,135.89) -- cycle ;
\draw  [color={rgb, 255:red, 245; green, 166; blue, 35 }  ,draw opacity=1 ][fill={rgb, 255:red, 255; green, 224; blue, 187 }  ,fill opacity=0.74 ] (516.13,133.69) -- (441.03,135.07) .. controls (448.93,134.92) and (441.7,121.48) .. (424.87,105.05) .. controls (408.04,88.62) and (387.99,75.41) .. (380.09,75.56) -- (455.19,74.18) .. controls (463.09,74.04) and (483.14,87.24) .. (499.97,103.67) .. controls (516.79,120.11) and (524.03,133.54) .. (516.13,133.69) -- cycle ;
\draw  [color={rgb, 255:red, 245; green, 166; blue, 35 }  ,draw opacity=1 ][fill={rgb, 255:red, 255; green, 224; blue, 187 }  ,fill opacity=0.67 ] (75.92,115.87) -- (177.06,113.1) -- (282.34,171.46) -- (181.2,174.23) -- cycle ;
\draw  [dash pattern={on 4.5pt off 4.5pt}]  (441.03,135.07) -- (510.5,174) ;
\draw  [dash pattern={on 4.5pt off 4.5pt}]  (517.22,133.66) -- (584.5,175) ;
\draw  [dash pattern={on 4.5pt off 4.5pt}]  (510.5,174) -- (584.5,175) ;
\draw    (296,147) -- (347.5,147.96) ;
\draw [shift={(349.5,148)}, rotate = 181.07] [color={rgb, 255:red, 0; green, 0; blue, 0 }  ][line width=0.75]    (10.93,-3.29) .. controls (6.95,-1.4) and (3.31,-0.3) .. (0,0) .. controls (3.31,0.3) and (6.95,1.4) .. (10.93,3.29)   ;
\draw [color={rgb, 255:red, 139; green, 6; blue, 24 }  ,draw opacity=1 ]   (230.2,142.28) -- (128.06,145.05) ;
\draw [color={rgb, 255:red, 139; green, 6; blue, 24 }  ,draw opacity=1 ]   (517.22,133.66) -- (443.24,135.89) ;

\draw (308,123.4) node [anchor=north west][inner sep=0.75pt]    {fold};
\draw (101,117.4) node [anchor=north west][inner sep=0.75pt]    {$M_{1}$};
\draw (393,98.4) node [anchor=north west][inner sep=0.75pt]    {$M_{1}$};
\draw (417,75.4) node [anchor=north west][inner sep=0.75pt]    {$\overline{M_{2}}$};
\draw (540,158.4) node [anchor=north west][inner sep=0.75pt]    {$\mathds{1}$};
\draw (175,155.4) node [anchor=north west][inner sep=0.75pt]    {$M_{2}$};
\draw (120,146.4) node [anchor=north west][inner sep=0.75pt]    {$I$};
\draw (432,139.4) node [anchor=north west][inner sep=0.75pt]    {$I$};

\end{tikzpicture}	
 	\caption{On folding the configuration of the membrane operators $M_1$ and $M_2$ on the left along the interface $I$, we find that there must be an interface between the membrane operator $M_1 \times \overline{M_2}$ and the trivial membrane operator $\mathds{1}$. Therefore, the fusion $M_1 \times \overline{M_2}$ must contain a condensation membrane operator.}
 	\label{fig: folding membrane operators}
 \end{figure}

Suppose $M_1$ and $M_2$ are two membrane operators in $\SC$ which admit a topological interface $I$ between them. Suppose $M_1$ is contained in the fusion 3-category $\SD_g$ and $\overline{M_2}$ is in $\SD_{h}$. Suppose $M_1$ and $M_2$ are connected along a topological interface $I$. On folding along the interface, we find that the membrane operator $M_1 \times \overline{M_2}$ must have an interface with the identity membrane operator (see Fig. \ref{fig: folding membrane operators}). Therefore, we have the fusion rule
\be
\label{eq:fusion of membranes and condensation operator}
M_{1} \times \overline{M_2}=C+\dots~,
\ee
 where $C$ is some condensation membrane operator and the $\dots$ represent other membrane operators in the fusion outcome. Since $C$ admits an interface with the identity membrane operator, it must be in the fusion 3-category $\SD_e$, where $e$ is the identity of the group $G$. To be consistent with this fusion rule, we must have $K_g \times K_h=K_e$. Therefore, $h=g^{-1}$ and $M_2 \in \SD_g$.  Moreover, for the fusion \eqref{eq:fusion of membranes and condensation operator} to be consistent with $K_g\times K_h=K_e$, its R.H.S must contain only membrane operators in $\SD_e=\SD$. Since $\SD$ describes the genuine topological operators of a 4+1d TQFT, the membrane operators and line operators in it must have a non-degenerate linking. Therefore, since $\Omega^2\SD\cong \text{Vec}$ by construction, $\SD$ does not contain any genuine non-condensation membrane operators. This shows that $M_1 \times \overline{M_2}$ must be a condensation membrane operator. This discussion shows that the set of equivalence classes of membrane operators $\pi_0(G)$, can be faithfully labelled by group elements $g\in G$. Moreover, the fusion of two membrane operators $M_1$ and $M_2$ in the equivalence classes labelled by $g$ and $h$, respectively, must result in a membrane operator in the equivalence class labelled by $gh$. Therefore, the set $\pi_0(G)$ can be equipped with a group-like fusion rule. Examples of fusion 3-categories with this group being $\DZ_2$ include 3Vec$_{\DZ_2}$ and Tambara-Yamagami 3-categories \cite{Choi:2021kmx,Kaidi:2021xfk,Bhardwaj:2024xcx}. Some fusion 3-categories with a non-abelian group structure on $\pi_0(G)$ arise as symmetries of class $\CS$ theories \cite{Bashmakov:2022uek,Antinucci:2022cdi}.
 
 Recall that the action of membrane operators on local operators only depends on the equivalence classes in $\pi_0(\SC)$. In particular, all membrane operators in the equivalence class containing the identity operator are condensation operators which act trivially on local operators. Since the representatives of the equivalence classes in $\pi_0(\SC)$ obey a $G$ fusion rule, the local operators are in a $G$-representation. Therefore, the action on local operators is invertible. 

\subsubsection{Dimensional reduction of topological membrane operators}

\label{sec: dimensional reduction of membrane operators}

Let $M$ be a membrane operator supported on $S^2 \times \mathds{R}$. On shrinking the $S^2$, we get some topological line operator $L_M$. Suppose the only topological line operator is the identity line operator, then we must have 
\be
L_M = n_M \trl ~ \forall M, 
\ee
where $n_M$ is some non-negative integer multiplicity. Using the same argument as in 2+1d, we find that if the membrane operator $M$ is simple, then the integer $n_M$ must be $1$. Therefore, all membrane operators on $S^2 \times \mathds{R}$ can be dimensionally reduced to the trivial line operator $L_M=\trl$. Now, consider the fusion rule
\be
M \times \overline{M} = C+ \dots
\ee
where $C$ is some condensation membrane operator. Dimensionally reducing both sides of the equality, we find a contradiction unless
\be
M \times \overline{M}=C~.
\ee
Consider the action of $M$ on some local operator $O$. Since the condensation operator $C$ acts trivially on $O$, we find that the action of $M$ must be invertible, with the inverse being the action of $\overline{M}$. Therefore, we find that the membrane operators act invertibly on local operators.

\subsection{Action of general finite non-invertible symmetries}

Consider a general finite non-invertible symmetry in 3+1d described by a fusion 3-category $\SC$. The topological line operators form $\Omega^2 \SC \cong \text{Rep}(H)$. On gauging this Rep$(H)$ 2-form symmetry, we get a symmetry described by a fusion 3-category $\SD$. We have $\Omega^2 \SD\cong \text{Vec}$. Therefore, from the results in the previous section, the equivalence classes of membrane operators $\pi_0(\SD)$ are faithfully labelled by a finite group $G$ and the action of membrane operators in $\SD$ on local operators is invertible. Consider the topological gauging interface $\CI$ between $\SC$ and $\SD$ obtained from gauging the Rep$(H)$ 2-form symmetry. Similar to our discussion on gauging interfaces in 2+1d, note that  $\CI^{\dagger} \times \CI$ is a condensation membrane operator in $\SC$ obtained from higher-gauging the 2-form symmetry Rep$(H)$ on a 3-manifold.

Let $M$ be a membrane operator in $\SC$. Consider its action on a local operator $O$. We have 
\be
M(O)= ((\CI^{\dagger}\times \CI) \times M \times (\CI^{\dagger} \times \CI))(O)= (\CI^{\dagger} \times M' \times \CI )(O)= (\CI^{\dagger} \times M')(O)~, 
\ee
where we have used the fact that $\CI$ acts by gauging the Rep$(H)$ 2-form symmetry, and hence does not act on local operators. Here $M'$ is a membrane operator in $\SD$ which is not necessarily a simple operator. Consider its decomposition
\be
M'=\sum_{U \in \SD} n_U U~,
\ee
where $n_U$ are non-negative integers, and $U$ is a simple membrane operator in $\SD$. The action of $M$ on $O$ can be written as 
\be
M(O)= (\CI^{\dagger} \times M')(O) = \CI^{\dagger}\bigg (\sum_{U \in \SD} n_U U(O) \bigg ) ~.
\ee
Note that the action of the simple membrane operators $U \in \SD$ on local operators is invertible since $\SD$ is a fusion 3-category without non-trivial line operators. This shows that the action of a general non-invertible symmetry $M$ on local operators can be decomposed into the action of some membrane operators $U$ which act invertibly on local operators, along with the action of the gauging interface $\CI^{\dagger}$. 

\label{sec: 3+1d action of non-inv with lines}

\subsection{Anomaly-free non-invertible symmetries in 3+1d}

\label{sec:anomaly-free 3+1d}

Consider the topological operators in a 3+1d QFT described by a fusion 3-category $\SC$. It is anomaly-free if it admits a trivially gapped phase. Mathematically, this should correspond to $\SC$ admitting an appropriate notion of a fibre 3-functor. In this section, we will assume that $\SC$ is anomaly-free and derive some consequences of this assumption. We will first consider non-invertible symmetries without topological line operators. We will show that all topological operators in $\SC$ can be rendered invertible through a topological manipulation. This shows that anomaly-free non-invertible symmetries in 3+1d without topological line operators are not intrinsically non-invertible. This generalizes the result for duality defects in 3+1d \cite{Sun:2023xxv}. Finally, we will study general anomaly-free non-invertible symmetries and derive some constraints on the topological line operators. 

The anomaly of duality symmetries in 3+1d was studied in \cite{
Apte:2022xtu,Cordova:2023bja,Antinucci:2023ezl}. 

\subsubsection{Non-invertible symmetries without topological line operators}

Consider a fusion 3-category $\SC$ describing a non-invertible symmetry in 3+1d implemented by topological membranes and surfaces. We will assume that the only topological line operator is the identity line operator. In this case, we learned in Section \ref{sec: 3+1d action of non-inv without lines} that one can define a group-like fusion rule on the components $\pi_0(\SC)$ given by a finite group $G$ and that the local operators of a QFT affording this symmetry are in a $G$-representation. Intuitively, one may guess that if this group $G$ has an anomaly, then this fusion 3-category must be anomalous. In this section, we will make this intuition precise by showing that the fusion 3-category $\SC$ can be converted to another fusion 3-category $\SD$ such that every connected component of $\SD$ contains at least one invertible membrane operator. We will show that these invertible membrane operators must implement a non-anomalous $G$ 0-form symmetry. 

Suppose the fusion 3-category $\SC$ is anomaly-free. Consider the 4+1d SymTFT described by the fusion 3-category $\CZ(\SC)$ and the canonical Lagrangian algebra $L_{\SC}$ corresponding to $\SC$. Since $\SC$ is anomaly-free, $\CZ(\SC)$ must admit a magnetic Lagrangian algebra $L_{\SE}$ where $\SE$ is the fusion 3-category of topological operators on the gapped-boundary $\CB_{\SE}$ defined by $L_{\SE}$. This algebra being magnetic to $L_{\SC}$ means that the operators which can end on $\CB_{\SC}$ cannot end on $\CB_{\SE}$, and vice-versa. A trivially gapped phase realizing the symmetry $\SC$ is obtained from interval compactification of the SymTFT with boundary conditions $\CB_{\SC}$ and $\CB_{\SE}$ \cite{Zhang:2023wlu,Putrov:2024uor}. The resulting phase must only have a single topological local operator. Therefore, line operators of the SymTFT, which are labelled by $\Omega^2\CZ(\SC)\cong \text{Rep}(G)$ which can end on the boundary  $\CB_{\SC}$ are not endable on the boundary $\CB_{\SE}$. Since $\SC$ does not contain any topological line operators, all line operators in the SymTFT $\CZ(\SC)$ must be endable on the boundary $\CB_{\SC}$. This implies that all non-trivial line operators in $\Omega^2\CZ(\SC)$ are not endable on $\CB_{\SE}$. Therefore, the bulk-to-boundary map $F_{\SE}:\CZ(\SC)\to \SE$ gives
\be
F_{\SE}(\Omega^2\CZ(\SC))\cong \text{Rep}(G)~.
\ee
 Note that, in general, the bulk-to-boundary map $F_{\SC}$ is expected to be surjective. This means that all simple topological operators in $\SC$ must appear in the image under the bulk-to-boundary map of some topological operator in the SymTFT $\CZ(\SC)$. However, when $F_{\SC}$ is applied to only topological line operators in $\CZ(\SC)$, it is not surjective in general. For example, if $\SE$ describes the symmetries of a TQFT, then the SymTFT $\CZ(\SE)$ does not contain any non-trivial genuine topological line operators \cite{Johnson-Freyd:2020usu}. This is because $\SE$ contains topological surface and line operators with non-degenerate linking. However, these operators cannot exist as genuine topological operators in the 4+1d SymTFT $\CZ(\SE)$ as topological surface and line operators braid trivially in 4+1d.\footnote{The line operators which braid non-trivially with surfaces arise from twisted-sector line operators in the SymTFT. For an explanation of this fact in the case of 3+1d SymTFT of 2+1d Chern-Simons theories, see \cite{Argurio:2024oym}.} 
 In this case, $F_{\SE}(\Omega^2\CZ(\SE))\cong \text{Vec}$ while $\Omega^2\SE$ is non-trivial for a non-trivial 3+1d TQFT. In general, $F_{\SE}(\Omega^2\CZ(\SE))$ is the subcategory of lines in $\SE$ which braids trivially with all surface operators. In our setting, $\SE$ is a fusion 3-category which is anomaly-free since $\CB_{\SC}$ is a magnetic boundary condition for $\CB_{\SE}$. Therefore, the braiding between surfaces and lines in $\SE$ must be trivial. Otherwise, the non-trivial braiding is a 1-form/2-form symmetry mixed anomaly which obstructs the existence of a trivially gapped phase. This discussion implies that
 \be
\Omega^2\SE= F_{\SE}(\text{Rep}(G))\cong \text{Rep}(G)~.
\ee
Let us define the fusion 3-category $\SD$ obtained from gauging the Rep$(G)$ 2-form symmetry of $\SE$. We have 
\be
\Omega^2 \SD\cong \text{Vec}~.
\ee
Moreover, since $\SD$ is obtained from gauging a symmetry of $\SE$, we must have
\be
\CZ(\SD)\cong \CZ(\SE)\cong \CZ(\SC)~. 
\ee
Therefore, $\Omega^2 \CZ(\SD)\cong \Omega^2(\CZ(\SC))\cong \text{Rep}(G)$. Applying the result in Section \ref{sec: 3+1d action of non-inv without lines} to $\SD$ we find that the Schur components $\pi_0(\SD)$ are labelled by the group $G$. Moreover, since $\SD$ is obtained from gauging a $\text{Rep}(G)$ 2-form symmetry of $\SE$, $\SD$ must contain a non-anomalous dual $G$ 0-form symmetry. In other words, $\SD$ contains a fusion 3-subcategory of the form 3Vec$_G$. This implies that the equivalence classes of membrane operators $\pi_0(\SD)$ each contain an invertible representative. 

The above discussion shows that there is a topological manipulation to go from the fusion 3-category $\SC$ to $\SD$ in which the membrane operators can be rendered invertible up to condensation defects. Therefore, an anomaly-free fusion 3-category $\SC$ without topological line operators is non-intrinsically non-invertible.  

\subsubsection{The general case}

\label{sec: anomaly-free general case}

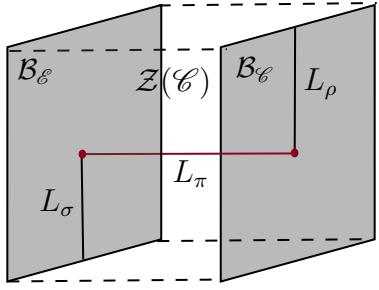
\begin{figure}[h!]
	\centering

\tikzset{every picture/.style={line width=0.75pt}} 

\begin{tikzpicture}[x=0.75pt,y=0.75pt,yscale=-1,xscale=1]

\draw  [color={rgb, 255:red, 0; green, 0; blue, 0 }  ,draw opacity=1 ][fill={rgb, 255:red, 74; green, 74; blue, 74 }  ,fill opacity=0.38 ] (335.15,210.1) -- (335.17,92.24) -- (412,71) -- (411.98,188.86) -- cycle ;
\draw [color={rgb, 255:red, 139; green, 6; blue, 24 }  ,draw opacity=1 ]   (267.5,146) -- (372.06,145.67) ;
\draw    (372.06,147.13) -- (372.29,82.14) ;
\draw  [dash pattern={on 4.5pt off 4.5pt}]  (229.64,94.14) -- (338.17,93.25) ;
\draw  [dash pattern={on 4.5pt off 4.5pt}]  (304.14,71.24) -- (413,70) ;
\draw  [dash pattern={on 4.5pt off 4.5pt}]  (303.45,189.76) -- (411.98,188.78) ;
\draw  [dash pattern={on 4.5pt off 4.5pt}]  (228.62,210.07) -- (335.15,209.1) ;
\draw  [color={rgb, 255:red, 139; green, 6; blue, 24 }  ,draw opacity=1 ][fill={rgb, 255:red, 139; green, 6; blue, 24 }  ,fill opacity=1 ] (370.65,145.4) .. controls (370.65,144.44) and (371.28,143.67) .. (372.06,143.67) .. controls (372.83,143.67) and (373.46,144.44) .. (373.46,145.4) .. controls (373.46,146.36) and (372.83,147.13) .. (372.06,147.13) .. controls (371.28,147.13) and (370.65,146.36) .. (370.65,145.4) -- cycle ;
\draw  [color={rgb, 255:red, 0; green, 0; blue, 0 }  ,draw opacity=1 ][fill={rgb, 255:red, 74; green, 74; blue, 74 }  ,fill opacity=0.38 ] (228.62,210.07) -- (228.65,92.21) -- (305.47,70.98) -- (305.45,188.83) -- cycle ;
\draw    (266.5,199) -- (265.69,146) ;
\draw  [color={rgb, 255:red, 139; green, 6; blue, 24 }  ,draw opacity=1 ][fill={rgb, 255:red, 139; green, 6; blue, 24 }  ,fill opacity=1 ] (264.69,146) .. controls (264.69,145.04) and (265.32,144.27) .. (266.09,144.27) .. controls (266.87,144.27) and (267.5,145.04) .. (267.5,146) .. controls (267.5,146.96) and (266.87,147.73) .. (266.09,147.73) .. controls (265.32,147.73) and (264.69,146.96) .. (264.69,146) -- cycle ;

\draw (341.17,95.65) node [anchor=north west][inner sep=0.75pt]  [font=\small]  {$\CB_{\SC}$};
\draw (309.5,147.55) node [anchor=north west][inner sep=0.75pt]    {$L_{\pi }$};
\draw (290.64,100.54) node [anchor=north west][inner sep=0.75pt]    {$\CZ(\SC)$};
\draw (375,101.4) node [anchor=north west][inner sep=0.75pt]    {$L_{\rho }$};
\draw (232.17,96.65) node [anchor=north west][inner sep=0.75pt]  [font=\small]  {$\CB_{\SE}$};
\draw (243,162.4) node [anchor=north west][inner sep=0.75pt]    {$L_{\sigma }$};

\end{tikzpicture}
	\caption{Consider the interval compactification of the SymTFT $\CZ(\SC)$ with gapped boundaries $\CB_{\SC}$ and $\CB_{\SE}$ to get a trivially gapped phase with $\SC$ symmetry. A bulk line operator $L_{\pi}$, where $\pi \in \text{Rep}(G)$ produces the line $L_{\rho}$ on the boundary $\CB_{\SC}$ where $\rho \in \text{Res}^G_H(\pi)$. Similarly, it produces the line operator $L_{\sigma}$ on the boundary $\CB_{\SE}$ where $\sigma \in \text{Res}^G_K(\pi)$. For a trivially gapped phase, the trivial line must not be in both $\text{Res}^G_H(\pi)$ and $\text{Res}^G_K(\pi)$ for any non-trivial $\pi$. }
	\label{fig: interval compactification}
\end{figure}

Consider a general fusion 3-category $\SC$ with 
\be
\Omega^2 \SC\cong \text{Rep}(H)~,
\ee
for some finite group $H$. Assume that $\SC$ is anomaly-free. The trivially gapped phase realizing $\SC$ can be obtained from the interval compactification of the SymTFT $\CZ(\SC)$ with boundary conditions $\CB_{\SC}$ and $\CB_{\SE}$, where $\CB_{\SE}$ is a magnetic boundary conditions with fusion 3-category of topological operators on it given by $\SE$. Let 
\be
\Omega^2 \SE\cong \text{Rep}(K)~,
\ee
for some finite group $K$. Moreover, let $\Omega^2 \CZ(\SC)\cong \text{Rep}(G)$.  Since $\CB_{\SE}$ is magnetic to $\CB_{\SC}$, a line operator $L_{\pi}$ for $\pi \in \text{Rep}(G)$ which can end on $\CB_{\SC}$ cannot end on $\CB_{\SE}$. Consider the bulk-to-boundary map $F_{\SC}: \CZ(\SC)\to \SC$ which describes how the operators in the bulk SymTFT get mapped to operators on the boundary condition $\CB_{\SC}$. By construction, acting on the topological line operators in the SymTFT, this map must be
\be
F_{\SC}(\text{Rep}(G))\cong \text{Rep}(H)~.
\ee
It must be surjective on the topological line operators, since $\SC$ is assumed to be anomaly-free. Explicitly, this map is given by $F_{\SC}(L_{\pi})=L_{\pi|_H}$ where $\pi|_H$ is the restriction of the representation $\pi \in \text{Rep}(G)$ to the subgroup $H$. Here surjectivity of the map $F_{\SC}$ follows from the fact that all irreducible representations of $H$ appear in $\pi|_{H}$ for some irreducible representation $\pi \in \text{Rep}(G)$. 
Similarly, for the boundary condition $\CB_{\SE}$, we find 
\be
F_{\SE}(\text{Rep}(G))\cong \text{Rep}(K)~,
\ee
which is again surjective on the topological line operators, since $\SE$ is also anomaly-free assuming $\SC$ is. This map is given by restricting the representations of $G$ to the subgroup $K$. Consider the line operator $L_{\rho}$ in $\SC$, where $\rho \in \text{Rep}(H)$. The set of line operators in the SymTFT, which on the boundary $\CB_{\SC}$ become a line operator containing $L_{\rho}$ are given by the set
\be
X_{H,\rho} :=\{L_{\pi}, \pi \in \text{Rep}(G) ~|~ \pi \in \text{Ind}_{H}^G(\rho) \}~.
\ee
Since the boundary $\CB_{\SE}$ is magnetic to $\CB_{\SC}$, the only line operator in $X_{H,\trl}$ which can end on $\CB_{\SE}$ is the trivial line operator (see Fig. \ref{fig: interval compactification}). Consider the line operator
\be
L=\sum_{\pi \in \text{Ind}_{H}^G(\trl) } n_{\pi}L_{\pi}~,
\ee
where $n_\pi$ is the multiplicity of the irreducible representation $\pi \in \text{Ind}_H^G(\trl)$. Fusing $L$ with the boundary $\CB_{\SE}$ must produce only one copy of the identity line operator in $\SE$. We have
\be
F_{\SE}(L)= \trl + \dots
\ee
Since $F_{\SE}$ acting on line operators is given by the restriction map, we have 
\be
\label{eq:condition on condensable lines}
\text{Res}_{K}^G L = \text{Res}_{K}^G\text{Ind}_{H}^G (\trl)  = \trl + \dots
\ee

The restriction of an induced representation can be explicitly determined using Mackey's restriction formula (for example, see \cite{serre1977linear})
\be
\text{Res}_{K}^G\text{Ind}_{H}^G (\trl)= \sum_{g \in K\backslash G/H} \text{Ind}_{{}^gH\cap K}^K ({}^g\trl)~,
\ee
where ${}^g H:=gHg^{-1}$ and ${}^g\trl = \trl$ is the conjugation action on the trivial representation. A representation induced from the trivial one always contains the trivial one in its decomposition into irreducible representations. Therefore for the formula above to be consistent with the requirement \eqref{eq:condition on condensable lines}, we must have 
\be
|K\backslash G /H|=1
\ee
For $g \in G$, the size of a double coset is given by 
\be
\label{eq:size of double coset}
|KgH|= \frac{|K||H|}{|K\cap {}^g H|}~.
\ee
Since double-cosets for different group elements are either equal or disjoint, for $K\backslash G /H$ to be the singleton set, we must have 
\be
|K e H|=|G|~.
\ee
This implies, 
\be
\label{eq: constraint on K and H 1}
\frac{|K||H|}{|K\cap H|}=|G|~.
\ee
Now, we will show that for the boundary conditions $\CB_{\SC}$ and $\CB_{\SE}$ to be magnetic to each other, the groups $H$ and $K$ must satisfy $H\cap K=\{e\}$ where $e$ is the identity element of the group $G$. To see this, note that the boundary condition $\CB_{\SC}$ is obtained from gauging a symmetry of the SymTFT $\CZ(\SC)$ given by the Lagrangian algebra $L_{\SC}$. The set of line operators that can end on the boundary is the set $X_{H,\trl}$, which contains line operators $L_{\pi}$ where $\pi \in \text{Rep}(G)$ such that $\pi|_H$ contains the trivial representation of $H$. These line operators form part of the symmetry $L_{\SC}$ that is gauged to produce the gapped boundary $\CB_{\SC}$. In 4+1d TQFTs, topological line operators braid non-trivially with 3d membrane operators. The connected components of membrane operators are labelled by conjugacy classes of the group $G$, say $M([g])$. The braiding of this membrane operator with a line operator $L_{\pi}$ is determined by the character $\chi_{\pi}([g])$. On gauging the lines in $X_{H,\trl}$, some of these membrane operators become purely confined (non-genuine) operators. In other words, they get attached to the boundary of 4d membrane operators implementing the dual 0-form symmetry. Let $[h]$ be a conjugacy class containing some $h\in H$, then the membrane operator $M([h])$ is not purely confined by gauging $X_{H,\trl}$ since $\pi|_{H}$ contains the trivial representation. For example, suppose $H$ is a normal subgroup of $G$, then $X_{H,\trl}$ is a fusion subcategory of Rep$(G)$ equivalent to Rep$(G/H)$. In this case, $\rho \in X_{H,\trl}$ when restricted to $H$ satisfies $\rho|_H=\text{dim}(\rho) \trl$ where $\trl$ is the trivial irreducible representation of $H$. Therefore, the membrane operators $M([h])$ for $h\in H$ braid trivially with all line operators in $X_{H,\trl}$ and they remain as genuine membrane operators after gauging $X_{H,\trl}$. More generally, when $H$ is not a normal subgroup, the line operators in the set $X_{H,\trl}$ are not closed under fusion. However, $X_{H,\trl}$ forms a gaugeable algebra object in Rep$(G)$ (for example, see \cite[Sec. 2]{kirillov2002q}). On gauging $X_{H,\trl}$, $M([h])$ becomes a direct sum of membrane operators in which some are genuine membrane operators while others are non-genuine. The presence of these genuine membrane operators in the image of $M([h])$ under gauging implies that, gauging $X_{H,\trl}$ does not confine all membrane operators and the resulting TQFT is not an SPT.\footnote{This phenomenon also happens in 2+1d TQFTs, where gauging a 1-form symmetry implemented by line operators which are not closed under fusion results in other line operators getting partially confined. See \cite[Section 6.3]{KNBalasubramanian:2025vum} for an example.}  However, gauging $L_{\SC}$ must result in an SPT so that it corresponds to creating a gapped boundary. Therefore, the membrane operator $M([h])$ must be part of the Lagrangian algebra $L_{\SC}$. In other words, it must be endable on the gapped boundary $\CB_{\SC}$. Similarly, the membrane operators $M([k])$ for $k\in K$ must be in the Lagrangian algebra $L_{\SE}$.  Therefore, for $L_{\SC}$ to be magnetic to $L_{\SE}$ we must have $H \cap {}^gK=\{e\}$ for all $g\in G$. Using \eqref{eq: constraint on K and H 1}, we find
\be
|G|=|K||H| \text{ and } |K \cap H|=1~.
\ee 
This shows that the group $G$ must be a Zappa-Sz\'ep product of the groups $H$ and $K$. The group $G$ is sometimes also called a bicrossed product $G=H\Join K$ of $H$ and $K$. In summary, we find the following result
\vspace{0.2cm}

\noindent \textit{A fusion 3-category $\SC$ with $\Omega^2 \SC\cong \text{Rep}(H)$ and $\Omega^2\CZ(\SC)\cong \text{Rep}(G)$ is anomaly-free only if there exists a subgroup $K\leq  G$} such that $G=H\Join K$. 
\vspace{0.2cm}

\noindent For coset non-invertible symmetries, this condition is shown in \cite{Hsin:2025ria}. It also holds for anomaly-free fusion 2-categories as shown in \cite{Decoppet:2023bay}. Our result shows that this is generally true for fusion 3-categories. Note that for a general non-invertible symmetry, the above condition is necessary but not sufficient for the symmetry to be anomaly-free. Indeed, for a fusion 3-category $\SC$ describing a 1-form symmetry $A$ with non-trivial anomaly, the 4+1d SymTFT $\DZ(\SC)$ is a 2-form gauge theory with gauge group $A$ \cite{Chen:2021xuc,Cordova:2023bja,Antinucci:2023ezl}. In this case, $G=\DZ_1$ and the above necessary condition is trivially satisfied. 

Consider an anomaly-free symmetry $\SC$ and an associated fusion 3-category $\SE$ obtained from the choice of a magnetic boundary condition for the SymTFT $\CZ(\SC)$. Let us define the fusion 3-category $\SD$ obtained from gauging the Rep$(K)$ 2-form symmetry of $\SE$. We have $\Omega^2 \SD\cong \text{Vec}$. Moreover, we must have
\be
\CZ(\SD)\cong \CZ(\SE)\cong \CZ(\SC)~. 
\ee
Therefore, $\Omega^2 \CZ(\SD)\cong \Omega^2\CZ(\SC)\cong \text{Rep}(G)$. Applying the result in Section \ref{sec: 3+1d action of non-inv without lines} to $\SD$ we find that the connected components $\pi_0(\SD)$ are labelled by the group $G$. Moreover, since $\SD$ is obtained from gauging a $\text{Rep}(K)$ 2-form symmetry of $\SE$, $\SD$ must contain a non-anomalous dual $K$ 0-form symmetry. In other words, $\SD$ contains a fusion sub-3-category of the form 3Vec$_K$. This implies that the connected components $\pi_0(\SD)$ labelled by $K\leq G$ each contain an invertible membrane operator. Note that, unlike the case considered in the previous section, here all connected components of $\SD$ do not necessarily contain an invertible membrane operator. 

Since the fusion 3-category $\SC$ contains topological line operators, the connected components $\pi_0(\SC)$ are not labelled by $G$, but rather by the double-coset $ H\backslash G/H$. To see this, note that we can gauge the Rep$(H)$ 2-form symmetry of $\SC$ to obtain a new fusion 3-category, say $\SN$ such that $\Omega^2 \SN \cong \text{Vec}$. Using the results in Section \ref{sec: 3+1d action of non-inv without lines}, we find that the connected components $\pi_0(\SN)$ are labelled by the group $G$. The fusion 3-category $\SC$ can be obtained from $\SN$ by gauging a dual 0-form $H$ symmetry implemented by membrane operators supported in $\pi_0(\SN)$ labelled by $H\leq G$. Under gauging such a symmetry, the membrane operators must be identified under the left and right action of the 0-form symmetry $H$. Therefore, the connected components of $\SC$ are labelled by the double coset $H\backslash G/H$. This double coset contains elements of the form $HgH$ for some $g\in G$. If $G=H\Join K$, then there exist unique $k\in K,\,h\in H$ such that $g=hk$. Therefore, $HgH=HhkH=HkH$. Therefore, the double cosets $H\backslash G/H$ are (not necessarily uniquely) labelled by the group elements $K$. Since $\SC$ can be converted to the fusion 3-category $\SD$ through a topological manipulation in which all connected components labelled by $K$ contain an invertible membrane operator, we find that the membrane operators in $\SC$ can be rendered invertible up to condensation in $\SD$.

\section{Conclusion}

\label{sec:conclusion}

We have argued that topological operators which admit a topological interface between them have identical action on local operators. For non-invertible symmetries in 2+1d and 3+1d, given by a collection of topological operators without non-trivial topological lines, we have shown that the action on local operators is invertible. More general non-invertible symmetries have an action which can be decomposed into an invertible action by some operators along with the action of a gauging interface. 

Let us note that our results also apply to the case of non-invertible symmetries with transparent fermionic line operators. For example, consider the dimensional reduction of a membrane operator $M$ on $S^2 \times \mathds{R}$, where $M$ is a simple object in a fusion 3-category $\SC$ such that the only non-trivial line operator is fermionic. Mathematically, we have $\Omega^2 \SC \cong \text{SVec}$. The line operator $L_M$ obtained from dimensional reduction must be bosonic. This is because the dimensional reduction of $M$ on $S^2 \times \mathds{R}$ can be understood as the action of $M$ on the trivial line operator $\trl$, and $\trl$ is a bosonic line. The action of a topological operator on a line operator cannot change its topological spin. Therefore, $L_M$ must be bosonic. From our assumption $\Omega^2 \SC \cong \text{SVec}$, we find that $L_M=\trl$ for all simple $M \in \SC$. Therefore, the consequences of this dimensional reduction, which are explored in Sec. \ref{sec: dimensional reduction of membrane operators} apply to fusion 3-categories with $\Omega^2 \SC \cong \text{SVec}$. 

Many of our results also generalize to higher-dimensions. In $d+1$ dimensions, we can consider a codimension one topological operator $M$ on the manifold $S^{d-1}\times \mathds{R}$. Shrinking the $S^{d-1}$ produces a line operator $L_M$. If there are no non-trivial topological line operators, then $L_M$ must be the trivial line $\trl$ for all $M$. This shows that the fusion rules of $M$ on $S^{d-1}\times \mathds{R}$ satisfies
\be
M \times \overline{M}=C~,
\ee
where $C$ is a condensation operator. Therefore, the action of $M$ on local operators must be invertible. Moreover, our result on anomaly-free non-invertible symmetries also generalize to higher dimensions. Indeed, the argument in Section \ref{sec: anomaly-free general case} can be repeated in $d+1$ dimensions to get the result:

\vspace{0.2cm}
\noindent \textit{A non-invertible symmetry described by a fusion d-category $\SC$ with $\Omega^{d-2} \SC\cong \text{Rep}(H)$ and $\Omega^{d-2}\CZ(\SC)\cong \text{Rep}(G)$ is anomaly-free only if there exists a subgroup $K\leq  G$} such that $G=H\Join K$.
\vspace{0.2cm}

There are several natural future directions to explore:
\begin{itemize}
	\item Our results leave room for an anomaly-free fusion 3-category symmetry which cannot be obtained from finite groups. It will be interesting to find an example of such a symmetry or to rule out this possibility by showing that anomaly-free non-invertible symmetries in 3+1d are all non-intrinsic. 
	\item We have shown that for non-invertible symmetries without topological line operators described by a fusion 3-category $\SC$, the Schur components $\pi_0(\SC)$ have a group structure given by a finite group $G$. It should be possible to reduce the anomaly data of the non-invertible symmetry given by the associators of the higher category to an effective anomaly of $G$. It will be interesting to get an explicit expression for such an anomaly. One promising route is it relate this effective anomaly to the 5-cocycle involved in the definition of the 4+1d SymTFT $\CZ(\CC)$. 
	\item While the action of non-invertible symmetries on local operators is very constrained, we are often interested in studying correlation functions involving non-gauge invariant point operators. These should be understood as point operators attached to appropriate Wilson lines so that the combined operator is gauge invariant. It would be very interesting to understand how non-invertible symmetries act on such operators and find consequences for their correlation functions.
	\item Topological operators in a TQFT, including non-invertible ones, play a crucial role in various quantum error correcting codes \cite{kitaev1997quantum,Hsin:2024pdi,Huang:2025ump,Radhakrishnan:2025gzl,Warman:2025hov}. It will be interesting to understand the consequences of our results in this context. 
	\item Symmetries on the lattice are forced to have a different structure than in the continuum. For example, there are some non-invertible symmetries described by a fusion category which can be realized on a lattice model with a tensor product Hilbert space only if it mixes with lattice translations \cite{Seiberg:2023cdc,Seiberg:2024gek,Evans:2025msy,Inamura:2026hif}. It would be interesting to understand how this influences the action of these symmetries on local operators of the lattice model. 
	\item This paper has focused on finite non-invertible symmetries. A natural open question is to understand the structure of continuous non-invertible symmetries \cite{Jia:2025vrj,Delmastro:2025ksn}. Various examples of such symmetries can be constructed from invertible continuous symmetries \cite{Antinucci:2022eat}. Moreover, there are also genuinely non-invertible continuous symmetries in 1+1d, which are not expected to be obtained from invertible symmetries \cite{Delmastro:2025ksn}. It would be very interesting to understand the structure of such symmetries in higher dimensions and see if they can have a richer action on local operators. 
\end{itemize}

\acknowledgments
We thank Mahesh Balasubramanian, Matthew Buican, Clement Delcamp, David Hofmeier, Ho Tat Lam, Jamie Pearson, Johann Quenta-Raygada, Apoorv Tiwari, Matthew Yu for related discussions. RR thanks Yunqin Zheng for several conversations related to this work, especially on coset symmetries. RR gratefully acknowledges the hospitality of University of California Los
Angeles, Kavli Institute for Theoretical Physics at UC Santa Barbara, Isaac Newton Institute for Mathematical Sciences, University of Birmingham, National Institute of Science Education and Research Bhubaneshwar, Indian Institute of Technology Bhubaneshwar, Indian Institute of Technology Madras, Indian Institute of Science Education and Research Thiruvananthapuram and the International Centre for Theoretical Sciences, where part of this work was carried out. The work of RR is supported by the UKRI Frontier Research Grant, underwriting the ERC
Advanced Grant ``Generalized Symmetries in Quantum
Field Theory and Quantum Gravity".

\appendix

\newpage
\bibliography{refs}

\end{document}